\documentclass[twoside,twocolumn,9pt]{article}
\usepackage{extsizes}

\usepackage[super,sort&compress,comma]{natbib} 
\usepackage[version=3]{mhchem}
\usepackage[left=1.5cm, right=1.5cm, top=1.785cm, bottom=2.0cm]{geometry}
\usepackage{balance}
\usepackage{mathptmx}
\usepackage{sectsty}
\usepackage{graphicx} 
\usepackage{lastpage}
\usepackage[format=plain,justification=justified,singlelinecheck=false,font={stretch=1.125,small,sf},labelfont=bf,labelsep=space]{caption}
\usepackage{float}
\usepackage{fancyhdr}
\usepackage{fnpos}
\usepackage[english]{babel}
\addto{\captionsenglish}{%
  
}
\usepackage{multicol}
\usepackage{makecell,multirow}

\usepackage{caption}
\usepackage{calc}
\usepackage{enumitem}
\newcommand{\tablistcommand}{
            \leavevmode\par\vspace{-\baselineskip}
                            }
\newcolumntype{P}[1]{p{#1-2\tabcolsep-\arrayrulewidth}}
\usepackage{colortbl}
\usepackage{array}
\usepackage{droidsans}
\usepackage{charter}
\usepackage[T1]{fontenc}
\usepackage[usenames,dvipsnames]{xcolor}
\usepackage{setspace}
\usepackage{float}
\usepackage[compact]{titlesec}
\usepackage{hyperref}

\usepackage{epstopdf}

\definecolor{cream}{RGB}{222,217,201}

\definecolor{Gray}{gray}{0.85}
\definecolor{LightCyan}{rgb}{0.88,1,1}

\begin{document}

\pagestyle{fancy}
\thispagestyle{plain}
\fancypagestyle{plain}{
\renewcommand{\headrulewidth}{0pt}
}

\newcommand{\articletitle}{Biomimicry in Nanotechnology: A Comprehensive Review}

\makeFNbottom
\makeatletter
\renewcommand\LARGE{\@setfontsize\LARGE{15pt}{17}}
\renewcommand\Large{\@setfontsize\Large{12pt}{14}}
\renewcommand\large{\@setfontsize\large{10pt}{12}}
\renewcommand\footnotesize{\@setfontsize\footnotesize{7pt}{10}}
\makeatother

\renewcommand{\thefootnote}{\fnsymbol{footnote}}
\renewcommand\footnoterule{\vspace*{1pt}%
\color{cream}\hrule width 3.5in height 0.4pt \color{black}\vspace*{5pt}} 
\setcounter{secnumdepth}{5}

\makeatletter 
\renewcommand\@biblabel[1]{#1}            
\renewcommand\@makefntext[1]%
{\noindent\makebox[0pt][r]{\@thefnmark\,}#1}
\makeatother 
\renewcommand{\figurename}{\small{Fig.}~}
\sectionfont{\sffamily\Large}
\subsectionfont{\normalsize}
\subsubsectionfont{\bf}
\setstretch{1.125} 
\setlength{\skip\footins}{0.8cm}
\setlength{\footnotesep}{0.25cm}
\setlength{\jot}{10pt}
\titlespacing*{\section}{0pt}{4pt}{4pt}
\titlespacing*{\subsection}{0pt}{15pt}{1pt}

\fancyfoot{}

\fancyfoot[LO,RE]{\vspace{-7.1pt}}
\fancyfoot[CO]{\vspace{-7.1pt}\hspace{20.2cm} \articletitle - \textit{Preprint Article} }
\fancyfoot[CE]{\vspace{-7.2pt}\hspace{4.2cm} \articletitle - \textit{Preprint Article} }

\fancyfoot[RO]{\footnotesize{\sffamily{ \thepage \hspace{2pt} of \hspace{2pt} \pageref{LastPage}}}}
\fancyfoot[LE]{\footnotesize{\sffamily{\thepage \hspace{2pt} of \hspace{2pt} \pageref{LastPage}}}}

\fancyhead{}
\renewcommand{\headrulewidth}{0pt} 
\renewcommand{\footrulewidth}{0pt}
\setlength{\arrayrulewidth}{1pt}
\setlength{\columnsep}{6.5mm}
\setlength\bibsep{1pt}

\makeatletter 
\newlength{\figrulesep} 
\setlength{\figrulesep}{0.5\textfloatsep} 

\newcommand{\topfigrule}{\vspace*{-1pt}%
\noindent{\color{cream}\rule[-\figrulesep]{\columnwidth}{1.5pt}} }

\newcommand{\botfigrule}{\vspace*{-2pt}%
\noindent{\color{cream}\rule[\figrulesep]{\columnwidth}{1.5pt}} }

\newcommand{\dblfigrule}{\vspace*{-1pt}%
\noindent{\color{cream}\rule[-\figrulesep]{\textwidth}{1.5pt}} }

\makeatother


\twocolumn[
  \begin{@twocolumnfalse}
{ \textit{Preprint Article}
}

\par
\vspace{1em}
\sffamily
\begin{tabular}{m{2.0cm} p{14.0cm} m{2.0cm} }

 & \noindent\LARGE{\textbf{\articletitle}} & \\
\vspace{0.3cm} & \vspace{0.3cm} \\

 & \noindent\large{Mehedi Hasan Himel\textit{$^{a, b}$}, Bejoy Sikder\textit{$^{a}$}, Tanvir Ahmed\textit{$^{a, b}$},  Sajid Muhaimin Choudhury\textit{$^{a}$} $^{\ast}$ } & \\

 & & \\
 &  \noindent\normalsize{Biomimicry has been utilized in many branches of science and engineering to develop devices for enhanced and better performance. The application of nanotechnology has made life easier in modern times. It has offered a way to manipulate matter and systems at the atomic level. As a result, the miniaturization of numerous devices has been possible. Of late, the integration of biomimicry with nanotechnology has shown promising results in the fields of medicine, robotics, sensors, photonics, etc. Biomimicry in nanotechnology has provided eco-friendly and green solutions to the energy problem and in textiles. This is a new research area that needs to be explored more thoroughly. This review illustrates the progress and innovations made in the field of nanotechnology with the integration of biomimicry. } &  \\

\end{tabular}

 \end{@twocolumnfalse} \vspace{0.6cm}

  ]

\renewcommand*\rmdefault{bch}\normalfont\upshape
\rmfamily
\section*{}
\vspace{-1cm}


\footnotetext{\textit{$^{a}$~Department of Electrical and Electronic Engineering, Bangladesh University of Engineering and Technology, Dhaka 1205, Bangladesh. }}

\footnotetext{\textit{$^{b}$~Department of Computer Science and Engineering, BRAC University, Mohakhali CA, Dhaka 1212, Bangladesh. }}

\footnotetext{$^{\ast}$ Corresponding Author: sajid@eee.buet.ac.bd }



\section{Introduction}
Nature, through billions of years of evolution, has often provided unique but elegant designs and structures that have inspired researchers and engineers to invent new technologies to enhance the capabilities of existing ones. The term 'Biomimetics' or 'Biomimicry', coined by American biophysicist Otto Schimt in the 1950s\cite{vincent2006biomimetics, schmitt1969some} is composed of two words that are derived from Greek words 'bios'(life) and 'mimesis'(imitation). Biomimicry is the emulation of natural structures, designs, and elements with a view to develop novel devices with the desired functionalities. \\

The evolving field of Biomimicry is highly multidisciplinary, and it is found on almost every engineering scale, from microscopic applications to macroscopic applications. The idea of mimicking nature to solve complex problems is implemented in every branch of science and engineering. Photonic structures are prevalent in nature such as the moth-eye, the structural coloring of insects, etc. provided many inspirations for the construction of novel photonic materials. Photosynthesis of plants and trees has provided the idea for newer energy harvesting methods. Mimicking the human brain's activity to solve problems such as object identification, pattern recognition, etc. has led to a new computation algorithm called 'Neural Network' which is now successfully implemented in many branches of science to solve complicated problems. Japan's "Shinkansen" or high-speed bullet train, when faced with the alarming problem of loud tunnel boom noise, was redesigned taking inspiration from nature: the front of the train was remodeled like a kingfisher’s face to achieve better aerodynamics. We are now able to make wind turbines with improved energy efficiency and performance, thanks to the "tubercles" (large bumps) design found in Humpback whales. Airplanes nowadays are designed almost like birds and advanced robotic shapes are looking more and more like humans. To push the limits of new innovations or to solve a complex engineering problem, humans have always looked at nature to seek answers throughout their extant. \\

Engineered biomimicry includes three methodologies: bioinspiration, biomimetics ,and bioreplication\cite{pulsifer2010mass}. Bioinspiration is the implementation of an idea taken from nature without reproducing the actual structure/mechanism. For example, both helicopters and bumblebees hover but the mechanisms are different. Biomimetics requires the reproduction of the actual mechanism to obtain a certain functionality. Robots that walk on four or more legs are instances of biomimetics. However, the distinction between these two is very slight, and differentiating them is not an easy task. The third approach bioreplication is the direct replication of a biological structure to obtain certain functionality. All of these methodologies exploit natural instances to develop structures or devices with desirable functionalities\cite{palma2013}. \\

As biomimicry enable researchers to exploit natural solution to solve a problem, nanotechnology provides the method for observing matter at a nanoscale level. Nanotechnology is the manipulation of matter at the nanometer $(10^{-9}m)$ scale. It is one of the most promising technologies of this century which has paved the way for many extraordinary innovations such as nano-sensors, nano-robots, etc. There is a close relationship between nanotechnology and biomimicry. All biological systems are comprised of units that are observed at the nanoscale and biology inspires the creation of new devices and systems to enhance the capabilities of the existing technology. Hence the integration of nanotechnology and biomimicry can be proved very essential in the amelioration of science and technology. \\

M Simovic-Pavlovic et al. have discussed the advancements of MEMS and NEMS in their work, along with designing new nanodevices for advanced materials and sensing applications\cite{simovic2022bioinspired}. PP Vijayan et al. have discussed the scope of biomimetic functionalities such as self-healing, self-cleaning, etc., for everyday materials\cite{vijayan2019biomimetic}. L Zhang et al. and ES Kim et al. have discussed the advancements of biomimetic tissue engineering and regenerative nanomedicine in their review works\cite{zhang2009nanotechnology},\cite{kim2014emerging}. T Li et al. and C Guido et al. have reviewed the recent progress of biomimetic cell membrane-coated nanoparticles and nanocarriers for cancer therapy\cite{li2020cell},\cite{guido2020biomimetic}. Z Jakšić et al. and A Giwa et al. have reviewed the development of biomimetic nanomembranes\cite{jakvsic2020biomimetic},\cite{giwa2017biomimetic} while R Bogue has reviewed biomimetic nano-adhesives\cite{bogue2008biomimetic}. N Soudi et al. have reviewed biomimetic photovoltaic solar cells\cite{soudi2020rise}, Y Saylan et al. have reviewed biomimetic molecular recognition and biosensing\cite{saylan2020advances}, and S Rauf et al. have reviewed biomimetic optical nanosensors\cite{rauf2016nano}. While many accomplished researchers and scientists of different fields have reviewed biomimicry in nanotechnology in specific areas, to our knowledge, there exists no comprehensive review of biomimicry in the general field of nanotechnology that encompasses different areas of the field in a single study. With that inspiration, in this work, we'll review various applications, advancements, and advantages of biomimicry in the field of nanotechnology.

A comprehensive overview of this review paper has been given in table \ref{tab:data_overview} which summarizes the application of biomimicry in various fields of nanotechnology.

\begin{table*}[hbt!]
\caption{Overview}
\label{tab:data_overview}
    \centering
\begin{tabular}{p{0.08\textwidth}p{0.18\textwidth}p{0.19\textwidth}p{0.2\textwidth}p{0.23\textwidth}}
    \hline
\thead{Fields} &   \thead{Applications} &   \thead{Biological Inspiration} & \thead{Significant Instances} & \thead{Features}    \\
    \hline
\multirowcell{16}{Nano-\\sensor}
           &   \multirowcell{7}{Optical \\ Nanosensor}  &
    \begin{itemize}[leftmargin=0cm,noitemsep,topsep=0pt,                     
                    before = \tablistcommand,
                    after  = \tablistcommand]
    \item Built-in IR sensing \newline
    capabilities of snakes and vampire bats\cite{shen2018bioinspired}
    \item Wide field-of-view of the\newline
    common housefly\cite{frost2015biomimetic}
    \item Camouflage capabilities\newline
    of cephalopods \cite{hanlon2007cephalopod}
    \end{itemize}      &      \begin{itemize}[leftmargin=0cm,noitemsep,topsep=0pt,                     
                    before = \tablistcommand,
                    after  = \tablistcommand]
    \item Heat sensors\cite{schmitz2012modelling}
    \hfill\break
    \hfill\break
    \item Aircraft wing deflection\newline
    measurement sensor\cite{frost2015biomimetic}
    \item Vapor sensors\cite{kertesz2018optical}
    \end{itemize}  &       \begin{itemize}[leftmargin=0cm,noitemsep,topsep=0pt,                     
                    before = \tablistcommand,
                    after  = \tablistcommand]
    \item Temperature sensing
    \item Chemical sensing
    \item Light-weight
    \item Low cost
    \item Better form factor
    \end{itemize}                  \\
   \cline{2-5}
   & \multirowcell{3}{Peptide \\ Nanosensor} & \begin{itemize}[leftmargin=0cm,noitemsep,topsep=0pt,                     
                    before = \tablistcommand,
                    after  = \tablistcommand]
    \item Peptide bonds
                    \end{itemize} & \begin{itemize}[leftmargin=0cm,noitemsep,topsep=0pt,                     
                    before = \tablistcommand,
                    after  = \tablistcommand]
    \item Nanoelectronic noses for\newline
    distinguishing chemical\newline
    vapors\cite{cui2012biomimetic}
    \end{itemize}      & \begin{itemize}[leftmargin=0cm,noitemsep,topsep=0pt,                     
                    before = \tablistcommand,
                    after  = \tablistcommand]
    \item Linking peptides to\newline
    nanosensors
    \end{itemize} \\
   \cline{2-5}
      &  \multirowcell{3}{Tactile \\ Nanosensor} &  \begin{itemize}[leftmargin=0cm,noitemsep,topsep=0pt,                     
                    before = \tablistcommand,
                    after  = \tablistcommand]
    \item Human skin\newline
    neuroreceptors
                    \end{itemize}  & \begin{itemize}[leftmargin=0cm,noitemsep,topsep=0pt,                     
                    before = \tablistcommand,
                    after  = \tablistcommand]
                    \item Skin attachable adhesives\newline
                    or E-skins\cite{park2014tactile}
                    \end{itemize} & \begin{itemize}[leftmargin=0cm,noitemsep,topsep=0pt,                     
                    before = \tablistcommand,
                    after  = \tablistcommand]
                    \item Reversible adhesion\cite{park2014tactile}
                    \item Ease of deformability\cite{park2014tactile}
                    \end{itemize}  \\
                    & & & & \\
   \cline{2-5} 
         & \multirowcell{4}{Stimuli \\ Responsive \\ Nanosensor} & \begin{itemize}[leftmargin=0cm,noitemsep,topsep=0pt,                     
                    before = \tablistcommand,
                    after  = \tablistcommand]
        \item Venus flytrap
                    \end{itemize}  & \begin{itemize}[leftmargin=0cm,noitemsep,topsep=0pt,                     
                    before = \tablistcommand,
                    after  = \tablistcommand]
                    \item Thermo-responsive \hfill\break
                    self-folding\newline
                    microgrippers\cite{breger2015self}
                    \end{itemize} & \begin{itemize}[leftmargin=0cm,noitemsep,topsep=0pt,                     
                    before = \tablistcommand,
                    after  = \tablistcommand]
                    \item Self-actuating action without\newline
                    any external power sources\cite{breger2015self}
                    \item Contains stimuli responsive\newline
                    hydrogel/polymers\cite{yoon2019advances}
                    \end{itemize}  \\
                    & & & & \\
    \hline
    \multirowcell{17}{Nano-\\medicine}
           &   \multirowcell{8}{Tissue Engineering}  &
    \begin{itemize}[leftmargin=0cm,noitemsep,topsep=0pt,                     
                    before = \tablistcommand,
                    after  = \tablistcommand]
    \item Human Bone Extracellular Matrix (ECM)
    \end{itemize}      &      \begin{itemize}[leftmargin=0cm,noitemsep,topsep=0pt,                     
                    before = \tablistcommand,
                    after  = \tablistcommand]
    \item Generation of bony tissues\cite{hartgerink2001self}
    \end{itemize}  &       \begin{itemize}[leftmargin=0cm,noitemsep,topsep=0pt,                     
                    before = \tablistcommand,
                    after  = \tablistcommand]
    \item Self-assembling \newline
    nanostructured gel\cite{hartgerink2001self}
    \end{itemize}                  \\ 
    & & \begin{itemize}[leftmargin=0cm,noitemsep,topsep=0pt,                     
                    before = \tablistcommand,
                    after  = \tablistcommand]
    \item Extracellular\newline
    matrix proteins\cite{hartgerink2001self,Matthews2002,Li2006,Wnek2003,Boland2001}
    \end{itemize}      &      \begin{itemize}[leftmargin=0cm,noitemsep,topsep=0pt,                     
                    before = \tablistcommand,
                    after  = \tablistcommand]
    \item Cell attachment, proliferation, differentiation and ultimately tissue generation.
    \end{itemize}  &       \begin{itemize}[leftmargin=0cm,noitemsep,topsep=0pt,                     
                    before = \tablistcommand,
                    after  = \tablistcommand]
    \item Creates scaffolds that mimic\newline
    natural tissues\cite{hartgerink2001self,Matthews2002,Li2006,Wnek2003,Boland2001}
    \end{itemize}                  \\
    & & \begin{itemize}[leftmargin=0cm,noitemsep,topsep=0pt,                     
                    before = \tablistcommand,
                    after  = \tablistcommand]
    \item Native Tissue ECM\cite{liu2017injectable}
    \end{itemize}      &      \begin{itemize}[leftmargin=0cm,noitemsep,topsep=0pt,                     
                    before = \tablistcommand,
                    after  = \tablistcommand]
    \item Hydrogels maintain  cell structure , repair  bond\cite{Yamaoka2006,liu2017injectable}
 
    \end{itemize}  &       \begin{itemize}[leftmargin=0cm,noitemsep,topsep=0pt,                     
                    before = \tablistcommand,
                    after  = \tablistcommand]
    \item Hydrogels provide strong\newline
    mechanical support\cite{Mauck2000} and\newline
    adjust themselves to match the shape of injury\cite{liu2017injectable} 
    \end{itemize}
    \\
    \cline{2-5}
    & \multirowcell{5}{Surgery}& \begin{itemize}[leftmargin=0cm,noitemsep,topsep=0pt,                     
                    before = \tablistcommand,
                    after  = \tablistcommand]
    \item Reconstruction of tissues
    \end{itemize}      &      \begin{itemize}[leftmargin=0cm,noitemsep,topsep=0pt,                     
                    before = \tablistcommand,
                    after  = \tablistcommand]
    \item Vascularized composite tissue allotransplantation (VCA)\cite{Amin2019}
    \item Tissue engineering
 
    \end{itemize}  &       \begin{itemize}[leftmargin=0cm,noitemsep,topsep=0pt,                     
                    before = \tablistcommand,
                    after  = \tablistcommand]
    \item Recovering tissues lost from\newline
    injury\cite{Amin2019}
    \end{itemize}
    \\
    \cline{2-5}

     &   \multirowcell{7}{Nanoparticles}  &
    \begin{itemize}[leftmargin=0cm,noitemsep,topsep=0pt,                     
                    before = \tablistcommand,
                    after  = \tablistcommand]
    \item Encapsulation of  
    \item[]dendritic cells\cite{Molino2013}
    \end{itemize}      &      \begin{itemize}[leftmargin=0cm,noitemsep,topsep=0pt,                     
                    before = \tablistcommand,
                    after  = \tablistcommand]
    \item Encapsulation of internal influenza proteins on the  VLPs\cite{Molino2013}
    \end{itemize}  &       \begin{itemize}[leftmargin=0cm,noitemsep,topsep=0pt,                     
                    before = \tablistcommand,
                    after  = \tablistcommand]
    \item Mimicking virus properties\newline
    increased vaccine effectiveness\cite{Patterson2013}
    \end{itemize}                  \\ 
    & & \begin{itemize}[leftmargin=0cm,noitemsep,topsep=0pt,                     
                    before = \tablistcommand,
                    after  = \tablistcommand]
    \item Nanocage
    \end{itemize}      &      \begin{itemize}[leftmargin=0cm,noitemsep,topsep=0pt,                     
                    before = \tablistcommand,
                    after  = \tablistcommand]
    \item Protein cage nanoparticle, encapsulin\cite{Lee2016}
    \end{itemize}  &       \begin{itemize}[leftmargin=0cm,noitemsep,topsep=0pt,                     
                    before = \tablistcommand,
                    after  = \tablistcommand]
    \item Protein based nanocages to\newline
    deliver drugs to specific sites\newline
    of human body\cite{Lee2016}
    \end{itemize}                  \\
    \\
    \hline
    \multirowcell{7}{Nano-\\robots}
           &   \multirowcell{3}{Bacteria Inspired  \\
           Nanorobots}  &
    \begin{itemize}[leftmargin=0cm,noitemsep,topsep=0pt,                     
                    before = \tablistcommand,
                    after  = \tablistcommand]
    \item Bacteria\cite{ali2017bacteria}
    \end{itemize}      &      \begin{itemize}[leftmargin=0cm,noitemsep,topsep=0pt,                     
                    before = \tablistcommand,
                    after  = \tablistcommand]
    \item Self-assembled flagellar \newline
    nanorobotic swimmers\cite{ali2017bacteria}
    \end{itemize}  &       \begin{itemize}[leftmargin=0cm,noitemsep,topsep=0pt,                     
                    before = \tablistcommand,
                    after  = \tablistcommand]
    \item Imitates the rotary motion of\newline
    helical bacteria flagella for\newline
    propulsion\cite{ali2017bacteria}
    \end{itemize}                  \\
   \cline{2-5}
   &   \multirowcell{3}{Platelet-Camouflaged \\ Nanorobots}  &
    \begin{itemize}[leftmargin=0cm,noitemsep,topsep=0pt,                     
                    before = \tablistcommand,
                    after  = \tablistcommand]
    \item Hybrid Red Blood\newline Cells\cite{esteban2018hybrid}, Platelets\cite{ali2017bacteria}
    \end{itemize}      &      \begin{itemize}[leftmargin=0cm,noitemsep,topsep=0pt,                     
                    before = \tablistcommand,
                    after  = \tablistcommand]
    \item Flagellar\newline nano-swimmers\cite{li2018biomimetic}
    \end{itemize}  &       \begin{itemize}[leftmargin=0cm,noitemsep,topsep=0pt,                     
                    before = \tablistcommand,
                    after  = \tablistcommand]
    \item Platelet imitating properties like binding to toxins or special kinds of pathogens\cite{esteban2018hybrid}
    
    \end{itemize}                  \\

\end{tabular}
    \end{table*}


\begin{table*}[h]
\caption*{Table 1 Overview(contd.)}
    \centering
\begin{tabular}{p{0.07\textwidth}p{0.18\textwidth}p{0.2\textwidth}p{0.2\textwidth}p{0.23\textwidth}}
    \hline
\thead{Fields} &   \thead{Applications} &   \thead{Biological Inspiration} & \thead{Significant Instances} & \thead{Features}    \\
    \hline

      \multirowcell{10}{Nano \\ Energy \\ Harvest}
           &   \multirowcell{8}{Solar Energy}  &
    \begin{itemize}[leftmargin=0cm,noitemsep,topsep=0pt,                     
                    before = \tablistcommand,
                    after  = \tablistcommand]
    \item Photosynthesis
    \item Plant Leaves\cite{chen2019biomimetic}
    \end{itemize}      &      \begin{itemize}[leftmargin=0cm,noitemsep,topsep=0pt,                     
                    before = \tablistcommand,
                    after  = \tablistcommand]
    \item Gratzel Cell\cite{smestad1998demonstrating,bhuvaneswari2013idea}
    \item Nano leaves and Nano trees\cite{kumar2014analyzing}
    \end{itemize}  &       \begin{itemize}[leftmargin=0cm,noitemsep,topsep=0pt,                     
                    before = \tablistcommand,
                    after  = \tablistcommand]
    \item Anthocyanin to recreate photosynthesisv\cite{smestad1998demonstrating,chen2019biomimetic}
    \item Thin photovoltaic film converts light into energy\cite{bhuvaneswari2013idea}
    \end{itemize}                  \\ 
    \cline{3-5}
    & & \begin{itemize}[leftmargin=0cm,noitemsep,topsep=0pt,                     
                    before = \tablistcommand,
                    after  = \tablistcommand]
    \item Butterfly Wings\cite{zhang2009novel}
    \item Compound Eye of insects\cite{chiadini2010simulation,martin2013engineered}
    \end{itemize}      &      \begin{itemize}[leftmargin=0cm,noitemsep,topsep=0pt,                     
                    before = \tablistcommand,
                    after  = \tablistcommand]
    \item Solar energy harvesting\cite{zhang2009novel,chiadini2010simulation}
    \end{itemize}  &       \begin{itemize}[leftmargin=0cm,noitemsep,topsep=0pt,                     
                    before = \tablistcommand,
                    after  = \tablistcommand]
    \item Increase of optical path length via light trapping\cite{zhang2009novel}
    \item Wide angular field and decreases the reflectance\cite{chiadini2010simulation,martin2013engineered}
    \end{itemize}                  \\ 
    \cline{2-5}
    & \multirowcell{2}{Ocean Energy} & \begin{itemize}[leftmargin=0cm,noitemsep,topsep=0pt,                     
                    before = \tablistcommand,
                    after  = \tablistcommand]
    \item Kelp
    \end{itemize} & \begin{itemize}[leftmargin=0cm,noitemsep,topsep=0pt,                     
                    before = \tablistcommand,
                    after  = \tablistcommand]
    \item Bio-inspired Triboelectric nanogenerator (BITENG)\cite{cui2018triboelectrification,he2018integrative,He2018, Choi, xu2018electron}
    \end{itemize}  & 
    \begin{itemize}[leftmargin=0cm,noitemsep,topsep=0pt,                     
                    before = \tablistcommand,
                    after  = \tablistcommand]
    \item Mimics the gentle sway of kelps\cite{Wang2019}
    \end{itemize} \\
    \hline
    
 \multirowcell{12}{Nano- \\ photonics}
           &   \multirowcell{7}{Anti-reflecting \\ Coating}  &
    \begin{itemize}[leftmargin=0cm,noitemsep,topsep=0pt,                     
                    before = \tablistcommand,
                    after  = \tablistcommand]
    \item Moth-Eye
    \end{itemize}      &      \begin{itemize}[leftmargin=0cm,noitemsep,topsep=0pt,                     
                    before = \tablistcommand,
                    after  = \tablistcommand]
    \item Integration of Moth-Eye nanostructures into solar cells \cite{Huang2013NanostructureAntireflection,forberich2008performance, zhu2010nanostructured, Yi2016BiomimeticStructures, yu2012recent, Chen2011BiomimeticCells, lan2021performance, ju2020fabrication, qarony2018approaching, tao2014tuning, sun2018biomimetic, kim2019moth, ji2013optimal, yun2020superhydrophobic}
    \item Superhydrophobic Surface in foldable displays \cite{yun2020superhydrophobic}
    \end{itemize}  &       \begin{itemize}[leftmargin=0cm,noitemsep,topsep=0pt,                     
                    before = \tablistcommand,
                    after  = \tablistcommand]
    \item Increase of External Quantum and Power conversion Efficiency of Solar cells
    \item Incident angle and polarization independent reflectance.
    \item Foldable displays exhibit excellent mechanical resilience with good thermal and chemical resistant. 
    \end{itemize}                  \\ 
    \cline{2-5}
    &   \multirowcell{4}{Anti-reflective \\ Transparent \\ Nanostrcutures}  &
    \begin{itemize}[leftmargin=0cm,noitemsep,topsep=0pt,                     
                    before = \tablistcommand,
                    after  = \tablistcommand]
    \item Transparent Wings 
    \item[] of \textit{Greta Oto}
    \item Cicada Cretensis \cite{binetti2009natural, siddique2015role, morikawa2016nanostructured}
    \end{itemize}      &      \begin{itemize}[leftmargin=0cm,noitemsep,topsep=0pt,                     
                    before = \tablistcommand,
                    after  = \tablistcommand]
    \item Omnidirectional 
    \item[] anti-reflective glass \cite{papadopoulos2019biomimetic} 
    \end{itemize}  &       \begin{itemize}[leftmargin=0cm,noitemsep,topsep=0pt,                     
                    before = \tablistcommand,
                    after  = \tablistcommand]
    \item Less than 1\% reflectivity irrespective of angle of incidence for S-P linearly polarized light.
    \end{itemize}                  \\
    & & & & \\
    \hline
    \multirowcell{8}{Textile \\ Engineering}
           &   \multirowcell{9}{Water Repellent Textile}  &
    \begin{itemize}[leftmargin=0cm,noitemsep,topsep=0pt,                     
                    before = \tablistcommand,
                    after  = \tablistcommand]
    \item Lotus Leaf\cite{marmur2004lotus,zhang2006superhydrophobic,balani2009hydrophobicity, ensikat2011superhydrophobicity, das2017potential}
    \end{itemize}      &      \begin{itemize}[leftmargin=0cm,noitemsep,topsep=0pt,                     
                    before = \tablistcommand,
                    after  = \tablistcommand]
    \item Superhydrophobic Surfaces on Cotton Textiles. \cite{shateri2013fabrication}
    \end{itemize}  &       \begin{itemize}[leftmargin=0cm,noitemsep,topsep=0pt,                     
                    before = \tablistcommand,
                    after  = \tablistcommand]
    \item Superhydrophobic, Antibacterial and UV-blocking fabric.
    \end{itemize}                  \\ 
    \cline{3-5} 
    & &
    \begin{itemize}[leftmargin=0cm,noitemsep,topsep=0pt,                     
                    before = \tablistcommand,
                    after  = \tablistcommand]
    \item Carnauba Palm (\textit{Copernicia Prunifera})
    \end{itemize}      &      \begin{itemize}[leftmargin=0cm,noitemsep,topsep=0pt,                     
                    before = \tablistcommand,
                    after  = \tablistcommand]
    \item Water repellent nano-coating on cotton, nylon and cotton-nylon fabrics.\cite{bashari2020bioinspired}
    \end{itemize}  &       \begin{itemize}[leftmargin=0cm,noitemsep,topsep=0pt,                     
                    before = \tablistcommand,
                    after  = \tablistcommand]
    \item Significant hydrophobicity during 30 sec and even after washing.
    \item Environment friendly water repellent textiles exhibiting good air permeability and antibacterial effect.
    \end{itemize}                  \\
    \cline{2-5}
    & \multirowcell{3}{Textile dyeing} & \begin{itemize}[leftmargin=0cm,noitemsep,topsep=0pt,                     
                    before = \tablistcommand,
                    after  = \tablistcommand]
    \item Prodigiosin\cite{liu2013mutant, falklof2016steric}
    \end{itemize}      &      \begin{itemize}[leftmargin=0cm,noitemsep,topsep=0pt,                     
                    before = \tablistcommand,
                    after  = \tablistcommand]
    \item Production of pyrrole 
    \item[] structure pigment by cell metabolization.\cite{gong2019bio}
    \end{itemize}  &       \begin{itemize}[leftmargin=0cm,noitemsep,topsep=0pt,                     
                    before = \tablistcommand,
                    after  = \tablistcommand]
    \item Better mass transfer rate for positive charged substances.
    \end{itemize}                  \\
\end{tabular}
    \end{table*}

\section{Biomimicry and Nanosensors}
Nanosensors are nano-devices used for sensing physical, real-life quantities that can later be converted into detectable signals. Although nanosensors are used in a variety of domains, they have a similar workflow: (i) binding, (ii) signal generation, and (iii) processing. Major advantages of using nanosensors (from their macro counterpart) are increased sensitivity and specificity, lower cost, and better response times. Taking inspiration from mother nature, researchers, and scientists are now trying to design nanosensors that very closely emulate nature’s sophisticated way of effortlessly binding to an analyte or generating signals. While there are many works that review nanosensors in light of specific applications\cite{munawar2019nanosensors}\cite{perdomo2021bio}\cite{srivastava2018nanosensors}\cite{vikesland2018nanosensors}, there still ceases to exist any comprehensive review on "biomimicry and nanosensors". In this section, we'll review some of the nanosensors which were designed friration from the biosphere, specifically in the following domains: optical nanosensing, peptide nanosensing, tactile nanosensing, and stimuli responsive nanosensing. 
\begin{figure*}
 \centering
 \includegraphics[width = \textwidth]{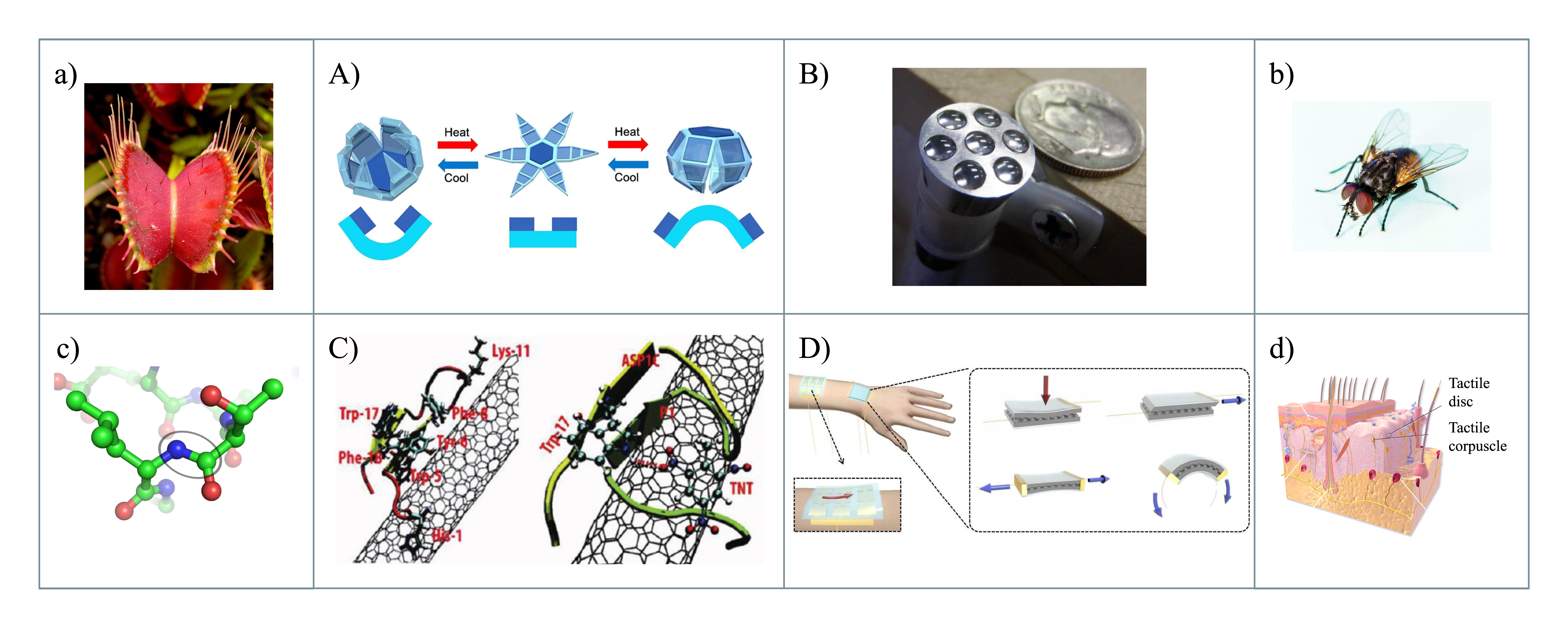}
 \caption{Biomimicry in nanosensors. a) Venus flytrap\cite{NoahElhardt} A) Thermo-responsive self-folding microgrippers\cite{breger2015self} [inspired by (a)] b) Common housefly\cite{USDAgov} B) aircraft wing deflection sensor\cite{frost2015biomimetic} [inspired by (b)] c) Peptide bonds\cite{webridge} C) Nano-electronic noses\cite{cui2012biomimetic} [inspired by (c)] d) Human skin neuroreceptors\cite{BruceBlaus} D) Skin attachable adhesives (E-skins)\cite{park2014tactile} [inspired by (d)]. (Images in this figure were reproduced from the cited original work, with permission from their corresponding publishers.)
}
 \label{fgr:example2col}
\end{figure*}
\subsection{Biomimetic Optical Nanosensors}
Research in the field of nanosensors have paved the way for the successful implementation of chemical, infrared and mechanical sensors, all of which have vast applications in modern-day technology, from tensile strength of wires to the design of an aerospace vehicle\cite{martin2019biomimetic}. \\
Many biological creatures have “built-in” infrared (IR) sensing capabilities. Emulating this particular characteristic has led to the fabrication of a biomimetic IR receptor micro-sensor. Some families of pythons, boas, and even beetles have infrared sensing organs, which sometimes convert the IR into heat, which in turn results in a thermal-sensitive system. In the inspired sensor, the incident IR radiation warms a liquid, thus inducing a mechanical deformation of the membrane, which is spotted in a capacitor\cite{herzog2015mu}. This is possible because temperature change can be detected by closely monitoring the changes in the capacitance. The main advantage of such a sensor is that it can be used at room temperature without any additional active cooling, unlike the other photo detectors. \\
Recently, another optical sensor that is inspired by the common housefly has been developed which can be used for aircraft wing deflection. The functionality of a compound eye is the wider field-of-view detection and other image processing techniques. The advantage of such a sensor is its low weight, less power consumption, faster computation, and better form factor.
Many animals use their ability to change their coloration for communication, predator capturing, camouflage, etc. This ability to change color has pushed engineers to make sensors that change color in presence of certain chemicals, hence acting as a chemical sensing sensor. The iridescent scales of some butterflies also show variation in their optical properties when exposed to different kinds of vapors\cite{potyrailo2007morpho}. These phenomena about the natural system inspired researchers to build photonic vapor sensors. Another device which consisted of periodic ridges and lamella deposited on a silicon wafer showed strong reflectance in well-defined spectral regions. These are some of the prime examples of how researchers, scientists and engineers have mimicked Mother Nature’s light exploitation techniques to fabricate optical nanosensors.

\subsection{Biomimetic Peptide Nanosensors}
By coupling peptide to a common nanomaterial surface (grapheme, carbon nanotube) a peptide nanosensor is created, which is capable of distinguishing chemically camouflaged mixtures of vapors and detecting chemical warfare agents to parts-per-billion level \cite{cui2012biomimetic}.

\subsection{Biomimetic Tactile Nanosensors}
Tactile wearable sensors that can receive information upon physical contact have gained a lot of attention from researchers and have potential usage in healthcare, robotic, and wearable device applications. Researchers are now working on tactile nanosensors that are based on biological tactile systems\cite{lee2020mimicking}. The main challenge of such a device is mimicking the human skin mechanoreceptors- receiving and transducing physical stimuli to electrical signals i.e. action potentials, and biomimetic signal processing mechanism\cite{lee2020biomimetic}.

\subsection{Biomimetic Stimuli Responsive Grippers}
Inspired by the Venus flytrap, many stimuli-responsive, biomimetic shape-changing robots/actuators have been proposed by combining stimuli-responsive hydrogels, polymer, or their hybrid combination\cite{yoon2019advances}. These grippers actually have far-reaching applications in the fields of biosensors, smart medicine, robotics, and surgery. Hydrogels and polymers are best suited for this purpose because they have the same moduli ranges ($\sim$KPa) of biological tissues and they can extend up to several orders of magnitude in volume when perturbed with external stimuli. This extending and shrinking mechanism mimics the self-actuating action without any external electrical power sources. The main material for this type of gripper is N-isopropylacrylamide (NIPAM) based stimuli-responsive hydrogels, and liquid crystalline material-based stimuli-responsive hydrogels.

\section{Biomimicry in Nano Medicine}
Biomimicry means technological achievements that have been developed using inspiration found in nature. The application of nanotechnology in healthcare is referred to as nanomedicine, and biomimicry is the inspiration for it because the human body is a gift of nature and riddled with questions and answers. In the sense of understanding natural structures and using this knowledge to perform and develop many tasks, nanomedicine may be regarded as a specialized division of biomimicry. Biomimicry assists in bone regeneration by use of self-assembled nanostructure tissues\cite{hartgerink2001self}, hydrogels\cite{Yamaoka2006,Mauck2000,Wu2018,liu2017injectable} or electrospinning\cite{martino2015extracellular}.Skin\cite{weyand2021biomimetics,girard2017biotechnological} and nerve restoration are another important aspect of biomimicry application in nanomedicine\cite{du2018biomimetic}. Biomimetic nanorobots are another application of biomimicry in nanomedicine. Inspired by platelets \cite{Olsson2005}, \cite{jbc2020} and platelet-membrane-cloaked nanomotors performs the task of the adsorption and isolation of platelet-targeted biological agents\cite{Li2018}. 
\subsection{Tissue Engineering and Surgery}

Plastic surgery is used to reconstruct tissues lost from injury, but it has evolved so much from the surgical practice of 30 years ago. Previously, doctors used synthetic implants using silicone\cite{Barr2011}, cellulose\cite{Barr2017} etc but they could interact with the host in a harmful manner, resulting in infection and rare tumors. And those implants did not mimic biological tissues. Vascularized composite tissue allotransplantation(VCA) was the first reconstructive milestone, but it had inevitable probabilities like transplant rejection. Tissue engineering then became the best alternative solution since this process does not have those limitations\cite{Amin2019}. Research of the self-assembling biomaterials that mimic complex biological structures has also been going on for a long time. Molecules that self-assemble into a nanostructured gel and mimic the key characteristics of human bone extracellular matrix (ECM). This is extremely important for the generation of bony tissues\cite{hartgerink2001self}. 

Biomimetic materials also support a microenvironment for cell attachment, proliferation, differentiation, and ultimately tissue regeneration. For example, the most abundant ECM protein in our body is collagen\cite{Ma2008}. A number of natural micromolecules ( collagen\cite{Matthews2002} , silk fibrin\cite{Li2006} , fibrinogen , synthetic polymers ( poly[glycolic acid]\cite{Wnek2003}, poly[L-lactic acid]\cite{Boland2001} , poly[lactic-co-glycolic acid] have been processed using electrospinning. This technique builds scaffolds that mimic natural tissues. To produce good scaffolds the biomaterial has to be both biocompatible and support cell growth. This will allow the scaffold to continue the metabolic functions. Moreover, It must promote integration with host tissue and allow vascularization\cite{martino2015extracellular}.  


Since the early 90s, multiple biomaterials have been exploited for cartilage and bone applications. And recently, hydrogels have become popular due to their network of interacting hydrophilic polymer chains that has striking similarities with native tissue ECM\cite{liu2017injectable}. Hydrogels maintain cell structure\cite{Yamaoka2006} and provide strong mechanical support\cite{Mauck2000}. In the case of bone repair, hydrogels bond with the natural ECM strongly\cite{Wu2018}. And hydrosols can be injected at the injury site, and they adjust themselves to match the shape of the injury\cite{liu2017injectable}. Artificial bone substitutes have been produced from mineral compositions, ceramic or material, bio-glass, or other fiber materials to mimic the organic tissue fraction as boney tissues\cite{brett2017biomimetics}. However, due to inability to fully master the level of growing fully functional bone with cortical bone components, bone replacement surgical techniques such as non-vascularized or vascularized bone transfer or the Masquelet technique to support bone construction by maintaining flap coverage are used\cite{brett2017biomimetics,buck2012bone,verboket2018autologous}.

Skin is also an important part of the human body that can receive damage from various sources, such as burns and infections. When a skin defect reaches the deep dermal layer, the self-regeneration potential of the skin is obstructed, and artificial skin substitutes using biomimicry can assist in skin repair\cite{weyand2021biomimetics}. Artificial skin substitutes have been constructed with different compositions and material properties, mimicking the organization of normal skin\cite{wood2014skin}. Examples are the collagen matrix bonded to flexible nylon fabric/silicone rubber epidarmis\cite{wood2014skin} or collagen-elastin matrix\cite{girard2017biotechnological}. Biomimetic scaffolds can also be used for nerve replacement by mimicking the inner structure and surface structure of original nerves\cite{du2018biomimetic}. 



\subsection{Biomimetic Nanorobots in Medical Science}

One exciting current application of biomimicry is the development of biomimetic nanorobots. These are micro-nanoscale robots for biomedical operations. Industrial robots are used mainly for the automation of dangerous tasks, but biomedical nanorobots are designed to tackle biological events. Biomimetic nanorobot design has recently started to take inspiration from nature, especially from the circulating cells such as platelets and leukocytes. Human platelets become the main source of inspiration owing to their work to tackle immune evasion\cite{Olsson2005}, \cite{jbc2020}, pathogen interactions, and hemostasis. The platelet membrane cloaking method is the wrapping of natural platelet cell membranes onto the surface of nanostructures and nanodevices\cite{Fang2017}. These platelet-membrane-cloaked nanomotors are enclosed by human platelets and do the adsorption and isolation of platelet-targeted biological agents\cite{Li2018}. These PL-motors possess a membrane coating containing multiple functional proteins related to platelets. These PL-motors evade the body’s immune system and display rapid locomotion in whole blood. PL-motors can also effectively absorb Shiga toxin using a Vero cell assay and also show binding to pathogens that can be used for rapid bacteria isolation\cite{Li2018}.


Minimally invasive surgery (MIS) and natural orifice transluminal endoscopic surgery (NOTES) is gradually taking the place of open surgery as this will decrease patient trauma and recovery length. Viewing the inside of a patient's lumen is a must for MIS. And now soft endoscopic nanorobots, inspired by nature, are being used for reaching distant surgical targets. Robot bodies are composed of soft materials that will reduce the potential of damaging tissue or organs. They are also bendable and flexible enough to move inside the patient's lumen. The Meshworm design developed by Bernth et al was inspired by the earthworm \cite{bernth2017novel}. The robot is made up of three segments of a soft plastic-silicone mesh composite. The total length of all segments is 50 cm and each segment is actuated by tendon wound mounted on DC motors, which are small enough to be embedded inside biomimetic meshworms. A PID controller controls the length of each tendon, and a USB camera is mounted at the tip.

\noindent Then there can be nanorobots that has hybrid cell membranes – combination of red blood cell (RBC) and platelet(PL) membranes. This shows better binding than PL-motors and neutralization of PL-adhering pathogens(Staphylococcus aureus bacteria) and neutralization of pore-forming toxins($\alpha$- toxin)\cite{DeAvila2018}. Some drugs can be carried by platelets because of the site-specific adhesion of platelets. PL - motors can also be used for this purpose\cite{Lu2019}.

\begin{figure*}[h]
    \centering
    \includegraphics[width = \textwidth]{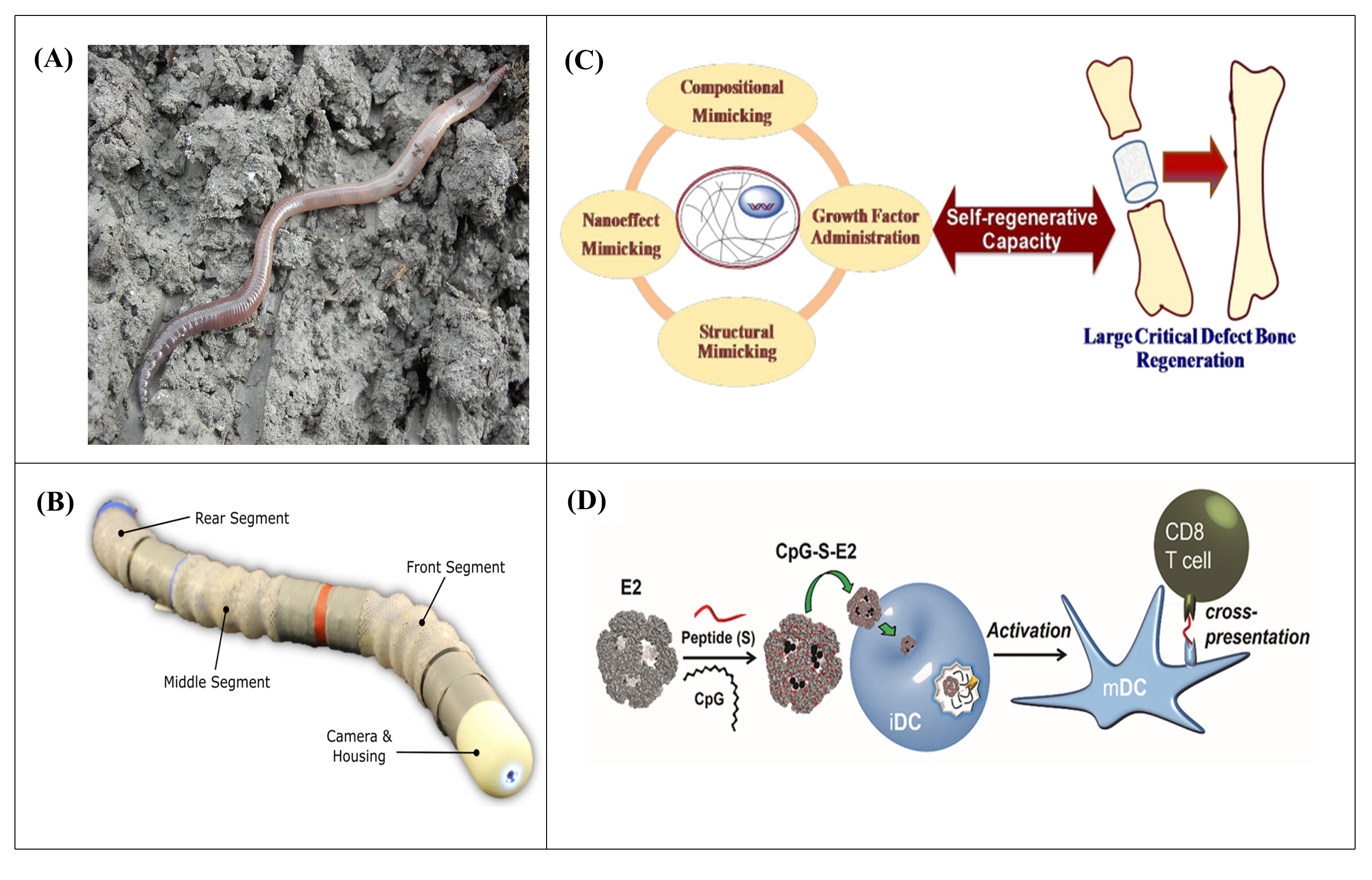}
     \caption{Biomimicry in Nanomedicine. A) Earthworm, B) A meshworm equipped with an endoscopic camera\cite{bernth2017novel} [inspired by (A), C) Schematic diagram of how a  combination of nanoelements and growth factor administration may allow regeneration of large critical defect bone\cite{li2017nanomaterial,khademhosseini2016decade}, D) Encapsulation of internal influenza proteins on the Virus-Like-Particles (VLPs)\cite{Molino2013} (Images in this figure were reproduced from the cited original work, with permission from their corresponding publishers.)}
    \label{fig_nano_medicine}
\end{figure*}

\subsection{Nanoparticles}
Nanoparticles(NPs) are usually defined as particles having a diameter of between 1 and 100 nm.  Nanoparticles have unique material characteristics owing to their submicroscopic size. For example, these nanoparticles can be used for the better activity of different medicines. Many cancer vaccines suffer from the inability to mount a T cell response. But viruses possess this size and the immune system can recognize these physical properties. So, mimicking these properties with nanoparticles produces a better platform for vaccine design. A viral mimicking vaccine platform capable of encapsulating dendritic cells showed a threefold better reaction than normal cases\cite{Molino2013}. A controlled immune response towards a specific virus-cell is also possible, which results in a vaccine that is active against the specific virus at a greater rate Encapsulation of internal influenza proteins on the interior of virus-like particles(VLPs) results in a vaccine that provides protection against 100 times lethal doses of an influenza\cite{Patterson2013}. Protein-based nanocages are also of particular interest to researchers because of their excellent ability to deliver a drug to specific sites of the human body. A protein cage nanoparticle, encapsulin, isolated from Thermotoga Maritima, has been developed for multifunctional delivery nanoplatforms for both chemical and genetic engineering\cite{Lee2016}. \\

Nanoparticles increase the bioavailability of compounds via their large surface-to-volume ratio. Many different nanoparticle systems have been created over the years to influence wound care in the medicine industry. Nanoparticles can also be tunable to react to external stimuli such as temperature\cite{zhao2016ph}, ultrasound\cite{kost1989ultrasound} or light\cite{afkhami2017entrococcus} which paves the way for targeted affinities toward certain tissues. Degradable nanoparticles such as polymeric NPs accelerate wound healing by controlling inflammation, fibroblast, and osteoclast activity\cite{baxter2013chitosan} and they can also be used as antibacterial agents and stem cells\cite{ahmed2015hydrogel}.

\section{Biomimicry and Nano-robots}
The recent research explosion in nanotech, with the new discoveries in the bio-molecular field, has brought about a new era in nanorobotics. Nanorobots of any nanostructure can currently handle the task of actuation, intelligence, information processing, sensing, etc. at the nano-scale. Some of the important functionalities of a nanorobot are (i) decentralized intelligence or “Swarm intelligence” (ii) self-assemble and replication at nano-scale (iii) signal processing at nano-scale (iv) nano to macro interface architecture\cite{ummat2005bionanorobotics}. In this section, we review some of the recent approaches in this field.
\begin{figure*}
 \centering
 \includegraphics[width = \textwidth]{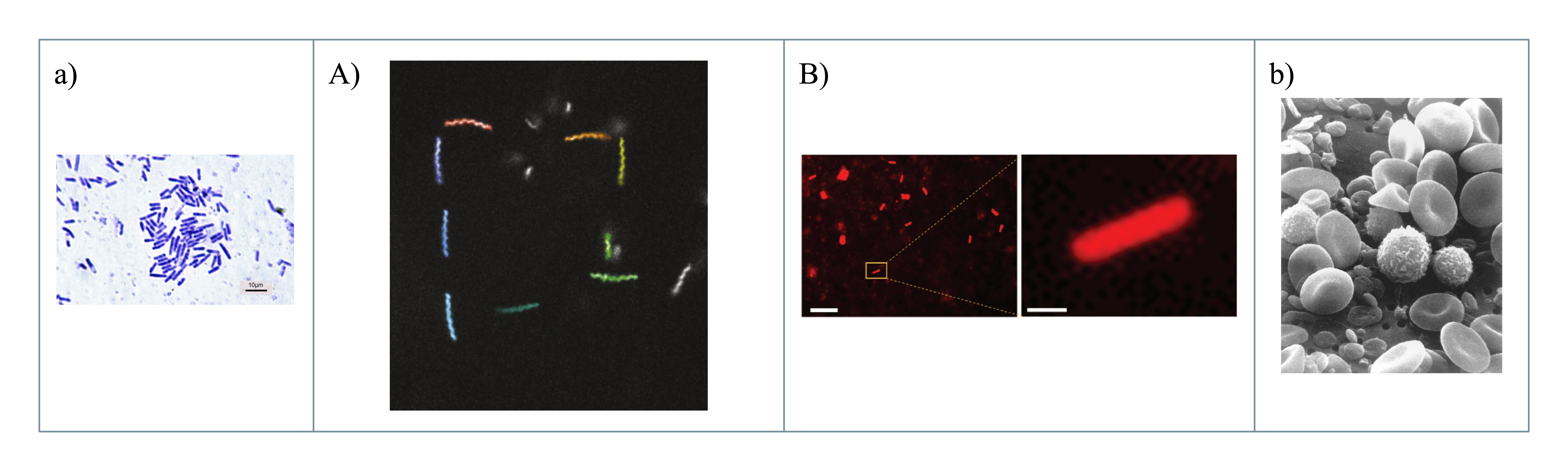}
 
 \caption{Biomimicry in nanorobots. a) Bacillus subtilis, a Gram-positive bacterium\cite{DrGrahamBeards} A) Bacteria inspired flagellar curly swimmer nanorobot\cite{ali2017bacteria} [inspired by (a)] b) SEM image from normal circulating human blood showing red blood cells, several white blood cells, and many small disc-shaped platelets\cite{wetzel1982scanning} B)  Fluorescent images of PL-motors covered with rhodamine-labeled platelet membranes. Scale bars, 20 µm (left) and 1 µm (right)\cite{li2018biomimetic} [inspired by (b)]. (Images in this figure were reproduced from the cited original work, with permission from their corresponding publishers.)
}
 \label{nano_robot}
\end{figure*}
\subsection{Biomimetic Bacteria Inspired Nanorobots}
Wireless nanorobots have had far-reaching biomedical applications. But so far most of the micro/nano swimmers have been imitating the rotary motion of helical bacteria flagella for propulsion, while none of them have the ability to reform itself in presence of different external stimuli. Recently, magnetic actuation of self-assembled flagellar nanorobotic swimmers has been discovered\cite{Ali2017}.
\subsection{Biomimetic platelet-camouflaged nanorobots}
While the main purpose of the industrial and life-sized robots is to work in a controlled manner, do some pre-determined tasks that are set beforehand, the task of biomedical nanorobots is to handle the different biological environments, adapt and improvise to the situation. In 2017, a study was published on a biologically interfaced nanorobot, which is made of magnetic helical nanomotors and is cloaked with the plasma membrane of human platelets \cite{Li2018}. This kind of nanorobots show platelet imitating properties, like binding to toxins or special kinds of pathogens. \\
A different but kind of similar study in 2018, reported the design, construction, and evaluation of a dual-cell membrane functionalized nanorobot for removal of biologically threat agents\cite{esteban2018hybrid}. In this study, the nanorobots consisted of gold nanowires and camouflaged in hybrid RBC and platelet membranes. This kind of nanorobots can fight against biological threats with no apparent biofouling while imitating natural cells. In the figure \ref{nano_robot}, the preparation and characterization of such a nanorobot are shown.
\section{Biomimicry and Nano Energy Harvest Technology}
The population of our beautiful world is expected to increase to 9 billion by 2050\cite{roberts20119}. This will result in an increase in the global energy consumption rate of 40.8 TW in 2050\cite{lewis2006powering}. But our fossil energy reserve is decreasing at an alarming rate, and it is extremely urgent to search for other alternative energy sources. Of these sources , solar energy has been the most researched, and the main focus is now to convert as much solar energy as possible to electrical energy.

\subsection{Solar Energy}
Energy is a must for humankind. The current focus of the world is on non-polluting sources of energy, and solar energy is the most desirable of them. 98\% of photovoltaic cells are silicon based\cite{osti_1174788} but solar cells require 99.999\% pure silicon, which is very energy intensive and its production steps also create hazardous byproducts. \\

Biomimicry is now being used to harvest solar energy from sunlight. The photosynthesis process used by plants leaves no waste. So the inspiration is there to use biomimicry in the solar cell. One such cell is the Gratzel cell \cite{smestad1998demonstrating} in which Gratzel used anthocyanin to recreate photosynthesis. This cell does not use silicon or other heavy metals, and this is much greener for the environment. The downside is that they are not as efficient as silicon cells.  \\

A Gratzel solar cell( also known as a dye-sensitized solar cell or DSSC) mimics the photosynthetic system by having $T_i O_2$ as   $NADP^+$ and $CO^2$, iodide as a water molecule, and -triiodide as oxygen. They harvest the solar energy together in nano-scale systems \cite{smestad1998demonstrating}. A basic DSC cell has an energy conversion efficiency of 1\% under direct radiation. But using thymol that has 10\% of the mass of the dye from cyanin 3-glucoside and cyanin 3-rutinoside improved the efficiency\cite{ali2007biomimicry}. The efficiency of DSSCs has exceeded 12\% after the last twenty years of development. The amount of light absorbed within the dye is an important part of determining efficiency. For this reason, suitable structures can improve the power conversion efficiency further\cite{hamann2008advancing},\cite{yella2011porphyrin}. \\ 

Another approach to improving the efficiency of DSC cells has been the modification and optimization of titania photoelectrodes. One approach was to use sphere  voids and large-sized solid particles as scattering center\cite{barbe1997nanocrystalline},\cite{hore2005scattering} or use multilayer structure\cite{wang2004significant} on the surface of the photoanode. These steps resulted in an increase in the optical path length of the photoanode and a subsequent increase in the light absorption efficiency. Another option is to use new structural materials that can increase the optical path length via light trapping.  Butterfly wings have microstructures that are effective solar collectors or blocks. Solar heat is absorbed at a faster rate at the ribs in the wing, which increases the body temperature faster. Then scientists discovered that honeycomb structures are less reactive than cross-ribbing structures, and that this structure also has a greater advantage in terms of light trapping. Scales have a high refractive index, and so total internal reflection occurs more, and nearly all incident light is absorbed. So, inspired by this, a butterfly wing scale titania film photoanode improved the light absorptivity of the DSC photoanode \cite{zhang2009novel}. This structure showed perfect light absorptivity and a higher surface area. This increased the total light harvesting efficiency and dye absorption. \\ 
\noindent Bioreplication has also become recently popular due to its application in solar-energy harvesting. Its application in solar energy is based on two observations, The first observation is the wide angular field that many insects, such as flies have. Each eye of a fly is a compound eye, consisting of multiple elementary eyes. These are arranged radially on a carved surface. The second observation is the almost halving of the reflectance, which is calculated by the simulation of a prismatic compound lens adhering to a silicon solar cell\cite{chiadini2010simulation}. \\
\noindent This experimental technique called the Nano4Bio technique\cite{martin2013engineered} has been designed to replicate the layer of a compound eye from the actual specimen. The idea is that by covering the surface of a solar cell with numerous replicas, the angular field of view of the solar cell will increase. The Nano4bio technique can produce multiple replicas simultaneously of multiple biotemplates. \\
\noindent Solar cell efficiency is mainly bound by light-harvesting efficiency, and some research is being done to use the photosynthetic leaf structures to increase the solar cell efficiency. Using nano-structured poly-carbonate thin-film inspired by the leaves of \textit{Salvinia cucullata} and \textit{Pistia stratiotes} helped to design a crystalline Si semiconductor with 18.1\% power conversion efficiency (PCE)\cite{yan2018photocurrent}. The light-trapping coating inspired by the epidermal cells in leaves helped in the fabrication of the graphene/Si Schottky junction\cite{das2019leaf}. The leaf anatomy combined upper and lower epidermis, palisade, and spongy mesophylls. This upper epidermal layer increases path length and allows light to traverse deep into the leaf. The middle layer has spongy mesophylls that scatter the light, and this scattered light has a higher chance of absorption by the lower part of the chlorophylls.

\subsection{Ocean Energy}
Ocean waves are a great source of renewable energy as they cover almost 70\% of the earth. But unlike solar and wind, ocean wave energy harnessing techniques have not developed so much yet\cite{schiermeier2016and, wang2016sustainably, wang2015triboelectric}. The integration of offshore wind power plants with ocean wave energy technology could enable power generation in a low-cost and more efficient way, which is obstructed by the deficit of matured wave energy harvesting technology\cite{sorensen2017commercial, wang2017new}. \\
\noindent To collect energy from wave energy, inspiration from kelp, a marine seaweed, has enabled the design of a triboelectric nanogenerator (TENG). Using the conjunction of triboelectric and electrostatic induction effects, TNG can convert mechanical energy into electrical energy\cite{cui2018triboelectrification,he2018integrative,He2018, Choi, xu2018electron}. Kelp-inspired TENG is made of vertically free-standing polymer strips which could sway independently to cause a contact separation with the neighboring strips, with the vibration of TENG in water mimicking the gentle sway of kelp. A single unit of BITENG (Bio-inspired TENG) provides an output short circuit current of about $10\mu A$ and an open circuit voltage of 260V with the maximum power density of $25\mu Wcm^{-2}$ which is high enough to drive 60 LEDs\cite{Wang2019}. A network of BITENG can collect more energy from ocean waves and be integrated with the offshore windmill for generating electricity.\\

\subsection{Nano Leaves and Nano Trees}
Another emerging application of biomimicry in the field of energy harvesting is nano leaves and stems of artificially created trees or plants\cite{kumar2014analyzing}. The nano leaves are distributed throughout artificial trees and plants, and they can supply a whole household's electricity demand in optimum condition\cite{save2015novel, Purohit, bogdanov2015role}. Nano leaf has two sides, one side has a very thin photovoltaic film that converts the light from the sun into energy. On the other side, thin thermo-voltaic films convert the heat from solar energy into electricity\cite{bhuvaneswari2013idea}. So, the sunlight then converts into electricity when it falls onto the nano leaves. Nano leaves convert both visible light and electrical light. When nano leaves feel strong winds, they can also convert wind energy to electrical energy. So nano leaves can create electricity in three ways\cite{Us}, directly convert solar energy to electricity but can also convert wind energy and solar heat to electricity.

\noindent Inspired by the leaf structure in nature, a biomimetic nanoleaf was developed for the first time\cite{chen2019biomimetic}. The fabricated nano leaves contain thin lamina and parallel veins, forming a monocot leaf structure. CuO nanowires with a high density on a conductive Cu mesh were used as the veins to support the layered double hydroxide(LDH) nanosheet lamina and promote charge transfer.
\begin{figure*}[h]
    \centering
    \includegraphics[width = \textwidth]{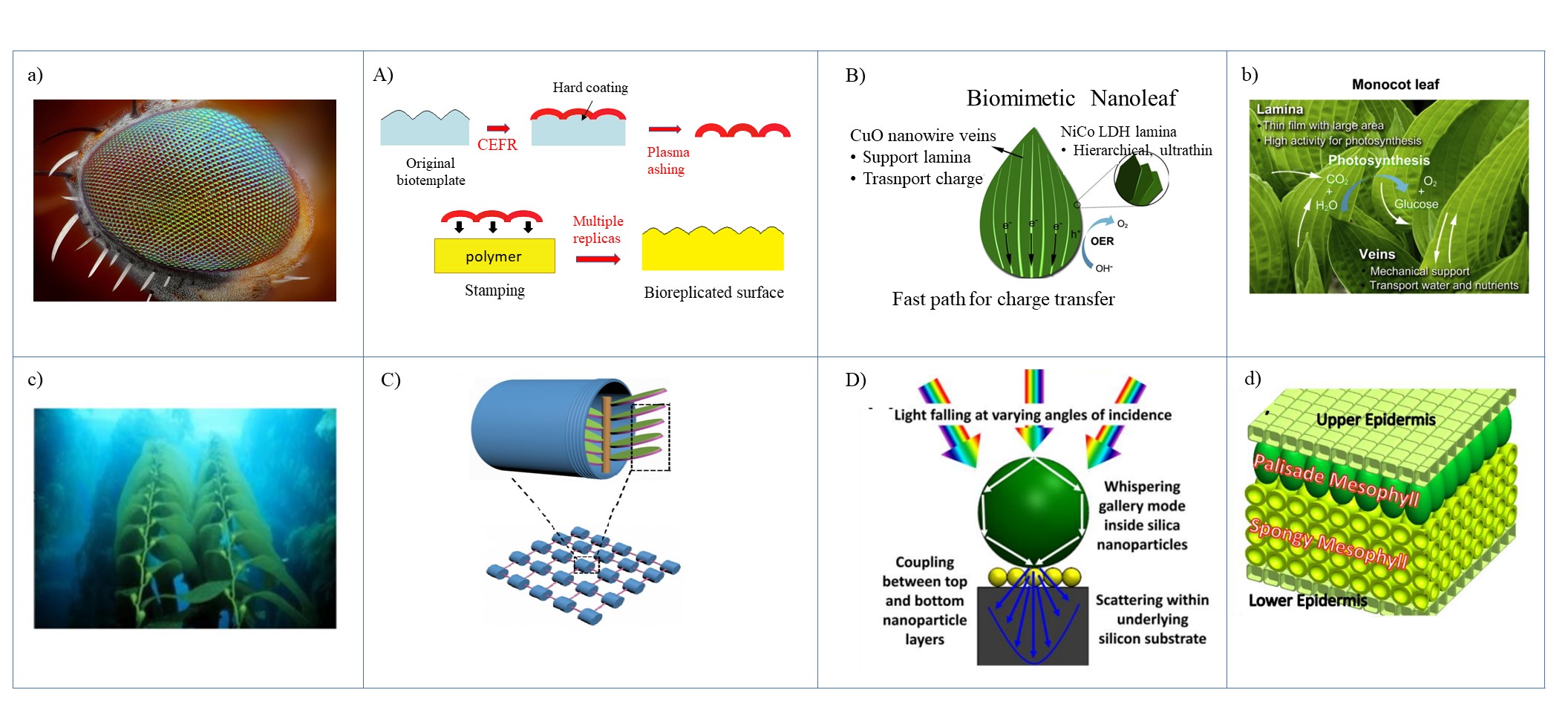}
     \caption{Biomimicry and nano energy harvest. a) Compound eye of fly, A) \textbf{Nano4Bio} technique developed to replicate the corneal layer of a compound eye from an actual specimen recreated from \cite{martin2013engineered} [inspired by (a)], b) Photograph of monocot leaf\cite{chen2019biomimetic}, B) Biomimetic nano leaf \cite{chen2019biomimetic} [inspired by (b)], c) Images of kelp plant \cite{Wang2019}, C) Schematic diagrams of a TENG-unit, the kelp-inspired TENG design and the networks\cite{Wang2019}[inspired by (c)],d) Schematic of leaf anatomy which inspired the light-trapping coating \cite{das2019leaf}, D) Light management mechanism in bilayer light-trapping scheme with all-dielectric sphere\cite{das2019leaf}[inspired by (d)]. (Images in this figure were reproduced from the cited original work, with permission from their corresponding publishers.)}
    \label{fig_nano_energy}
\end{figure*}


\section{Biomimetic Nanostructures and Photonics}
\begin{figure*}[h]
    \centering
    \includegraphics[width = \textwidth]{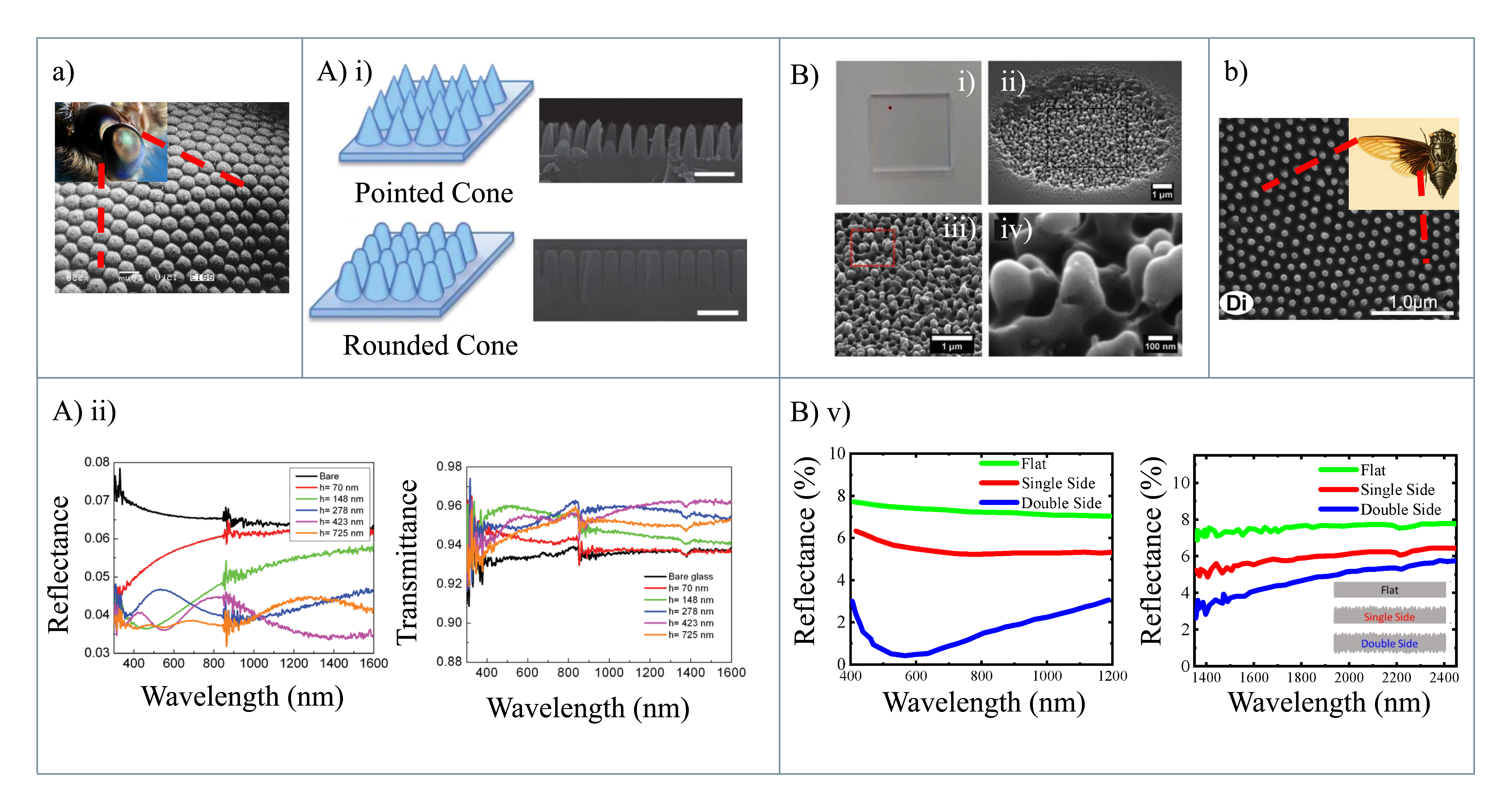}
    \caption{Biomimetic structures in Nanophotonics. a) Actual and SEM images of the eyes of a moth\cite{RickCowen}\cite{ShirleyJWright} A) i) Moth eye inspired anti-reflective nano cones. Upper one is the pointed cone and its SEM image. Lower one is the rounded cone and its SEM image. The scale bars in the SEM images are all 500 nm [inspired by (a)] ii) Measured reflectance and transmittance of the moth eye inspired nano cones and the bare structures\cite{ji2012improved} [inspired by (a)] b) SEM image of Cryptotympana atrata [cicada species] wing surfaces\cite{WLDistant}\cite{sun2012influence} B) i-iv) Photographs and SEM image of an omnidirectional anti-reflective glass inspired from Cicada Cretensis scale\cite{papadopoulos2019biomimetic} [inspired by (b)] v) Reflectance sprectra of a pristine (green lines) and laser treated at one (red lines) or both sides (blue lines) fused silica plate \cite{papadopoulos2019biomimetic}. Reflectance in both cases decreases with the introduction of biomimetic anti-reflection coating [inspired by (b)](Images in this figure were reproduced from the cited original work, with permission from their corresponding publishers.)}
    \label{photon1}
\end{figure*}
Over the years, nature, through evolution, has inspired photonics researchers by providing a great variety of biological nanostructures. This has enabled them to design new devices and modify the existing ones to enhance performance. Prevalent biological photonic structures are developed and optimized to perform necessary tasks for that creature, such as camouflage, courtship, warning or attractant to conspecifics, pollination, and so on\cite{Parker2007BiomimeticsNanostructures, Vukusic2003PhotonicSystems,Xu-2013}. Structural coloration and iridescence of insects, birds' feathers, plants \cite{ghiradella1991light, Vukusic2003PhotonicSystems, prum2003structural, stavenga2011dramatic, vignolini2013analysing}, densely arrayed triangular hairs on Saharan ants' for thermoregulation \cite{shi2015keeping}], anti-reflecting coating of the moth and butterfly eyes \cite{wilson1982optical, Bernhard1972} and on the transparent wings of hawkmoth\cite{yoshida1997antireflective}, the apparent 'metallic' view of fish-scale \cite{Parker2007BiomimeticsNanostructures}etc. are the instances of photonic structures existent in nature. In this section, we'll review the applications of such natural structures in the area of nanophotonics. Mainly we'll review how biomimetic structures can improve the absorption, reflectance, and transmittance spectra. Figure \ref{photon1} summarizes the applications of a few such biomimetic structures in nanophotonics. \\

In 1967, Bernhard discovered that the reflection over the visible range (400 to 800 nm) from the corneas of night flying moths is reduced because of their having tiny conical burls \cite{bernhard1967structural}, \cite{clapham1973reduction}. This has led to the discovery of the artificial moth eye, which is a very fine array of protuberances. The working of an artificial moth-eye surface may be understood with the help of the surface layer, which has a gradually changing refractive index having a value from unity to that of the bulk. Due to the presence of gradually varying refractive index layers, the net reflectance is the resultant of an infinite series of reflections at each incremental change in the index. Each of these reflections has different phase factors because of the change in depth of different layers. Now if this occurs over a distance of $\frac{\lambda}{2}$, all the phases will be present, which will result in destructive interference and so the net reflectance will be zero. \\

\subsection{Moth-Eye Nanostructures}
\begin{figure*}[t]
    \centering
    \includegraphics[width = \textwidth]{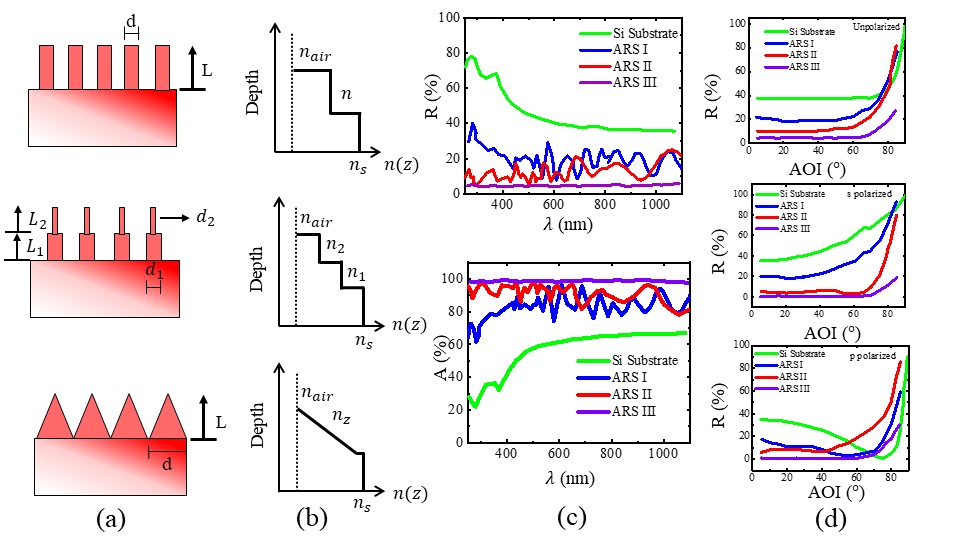}
 \caption{(a) Schematic diagram of three types of Si ARS with corresponding geometrical parameters. (b) The refractive index profile for each ARS. The refractive index of air, Si and mixture of air/Si is represented by $n_{air}$, $n_s$, $n_{1/2}$ respectively. $n_z$ represents the gradient index of refraction as a function of depth, which is in the z direction. (c) Reflectance and absorption as a function of wavelength for three types of ARS with a reference planar Si substrate. R stands for reflectance and A stands for absorption. (d) Reflectance data for Si wayfer, ARS I, ARS II, ARS III as a function of angle of incidence (AOI) for unpolarized light,  s-polarized light, and p-polarized light of 633nm wavelength light. The spectrum is almost independent of incident angle for biomimetic ARS.}
    \label{photon2}
\end{figure*}

The spacing of the protuberances of the moth-eye must be small enough that the array cannot be resolved by the incident light; otherwise, the array will work as a diffraction grating and the light will simply be redistributed into the diffracted orders\cite{wilson1982optical}. The optical property of an artificial moth-eye surface depends on the height of the protuberances and their spacing. The pattern of the moth-eye array is the photo-resist interference fringe at the intersection of two coherent laser beams\cite{clapham1973reduction}. The advantage of moth eye anti-reflection coating applied to a high-quality optical component comes at larger angles of incidence. In the case of normal incidence, the moth-eye has the same performance as conventional multilayer coating techniques. But at a larger angle of incidence performance of the former is better than the latter. In a work presented in 2013, a group showed the comparative study of three different anti-reflecting structures (ARS) for broadband and angle-independent anti-reflection\cite{Huang2013NanostructureAntireflection} both theoretically and experimentally. These structures were: single diameter nanorods (ARS I), dual diameter nanorods (ARS II), and biomimetic nanotips (ARS III) which take after the submicron structure of moth'e eyes. The simulations were done using finite difference time domain calculations. All the structures were made of Si. The structures along with the refractive index profile are shown in figure \ref{photon2} (a)-(b).\\
\noindent The three structures, along with a planar Si wafer (used as reference) were simulated using FDTD to obtain the results of reflectance and absorption over 250-1100nm. Figure \ref{photon2}(c) shows the resulting reflectance and absorption over the mentioned wavelength region. From figure \ref{photon2}(c) it is seen that ARS-III has the lowest reflectance value among the others and in the case of absorption shown in figure \ref{photon2}(b), it has almost 98 \% absorption over the wavelength range. \\

\noindent The dependence of reflectance on the angle of incidence (AOI) was studied from 5$^\circ$ to 85$^\circ$ using unpolarized, s-polarized, and p-polarized light of 633nm wavelength. It is seen from figure \ref{photon2}(d) that irrespective of polarization, the reflectance of ARS III structure shows almost no AOI dependence below brewster angle ($\sim75^\circ$). From these results, it is concluded that ARS III inspired by the structure of moth eyes, has shown better performance as anti-reflector than both ARS I and ARS II. \\

In the case of the solar cell, the greater the absorption of the incident light, the greater the efficiency. So moth-eye is integrated into solar cells to increase the efficiency of solar cells. As we have seen earlier, moth-eye anti-reflecting coating has virtually zero dependence of reflectance on the angle of incidence. This property can be exploited in an organic solar cell to enhance performances\cite{forberich2008performance}. Forberich et al \cite{forberich2008performance} fabricated a type of organic solar cell with a moth-eye anti-reflective coating as an effective medium at the air-substrate interface.

The organic solar cell consisted of a thin ($\sim$ 20 nm) layer of poly
(3,4-ethylenedioxythiophene) doped with poly(styrene sulfonate) (PEDOT:PSS Baytron, Bayer AG) on indium tin oxide (ITO) coated on a glass substrate. The moth-eye was introduced at the air-glass interface.

\begin{figure}[h]
    \centering
    \includegraphics[width = 0.5\textwidth]{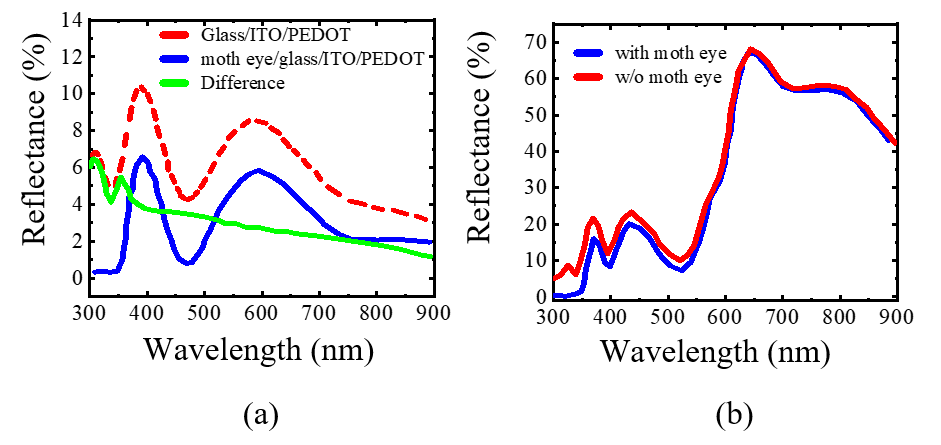}
    \caption{Measured reflection with and without moth eye from a layer stack of (a) glass/ITO/PEDOT and (b) P3HT:PCBM solar cell.\cite{forberich2008performance} }
    \label{photon3}
\end{figure}

\noindent The measurement results are shown in figure \ref{photon3}(a) and (b). From the reflectance vs wavelength curve for glass/ITO/PEDOT stacks, both curves show oscillatory behavior due to interference in the dielectric slab. However, in the case of moth eye/glass/ITO/PEDOT, the overall reflection is reduced due to the introduction of the moth-eye layer. Figure \ref{photon3}(b) shows the result of state-of-the-art P3HT/PCBM solar cells with and without moth eyes. \\

Here it is observed that the highest reduction of reflection occurs in the range of 350 to 600 nm wavelength range. And beyond that, the reflection becomes almost identical. This happens because, in the case of longer wavelengths, most of the incident light is reflected from the metal electrode, and so the reduction of reflection at the air-glass interface has no significant effect on the overall reflection. From the measured incident angle dependence of external quantum efficiency (EQE) of the P3HT:PCBM solar cell with a moth-eye anti-reflective coating,  it is found that at 0$^\circ$ incident i.e., normal incident, the EQE increases by around 3.5\% with the introduction of moth eyes in a complete solar cell. In fact, EQE is greater with moth eyes than with the normal solar cell for all angles of incidence. \\

Zhu et al \cite{zhu2010nanostructured} reported a nanocone solar cell. It had an efficient optical property with a great processing advantage over nanowire and thin-film solar cells. The work found that the nanocone arrays had shown absorption above 93\% which was much greater than the nanowire arrays (75\%) or thin films(64\%) in the range of 400-600 nm of wavelength\cite{Yi2016BiomimeticStructures},\cite{yu2012recent}.\\

In another work, biomimetic nanostructured antireflection coatings with polymethyl methacrylate (PMMA) layers were integrated with silicon crystalline solar cells. The structure could reduce the average reflectance of solar cells from 13.2\% to 7.8\%. As a result, power conversion efficiency increased from 12.85\% to 14.2\%\cite{Chen2011BiomimeticCells}.\\

The anti-reflector moth eye used in perovskite solar cells can improve performance significantly. All-inorganic carbon-based perovskite solar cells (PSCs) have drawn increasing interest among researchers due to their low cost and the balance between bandgap and stability\cite{lan2021performance}. However, due to the reflection at the air/glass interface, such PSCs suffer an optical loss of about 10\% of the incident photons\cite{rad2020anti, sertel2019effect}. This optical loss results in a reduction in the short circuit current density $(J_{sc})$ for PSCs. The inclusion of moth-eye antireflection (AR) nanostructures with wide-range-wavelength AR property has paved the way to solve the optical loss issue of the above-mentioned PSCs and enhance the power conversion efficiency (PCE)\cite{lan2021performance, ju2020fabrication, qarony2018approaching, tao2014tuning, sun2018biomimetic, kim2019moth, ji2013optimal}. Compared to the conventional AR, the additional benefit of moth-eye ARs is the enhancement of broadband antireflection through the tuning of structural and geometrical parameters i.e. height, periodic distance, shape, and arrangement\cite{ji2013optimal}. Such modulation of optical properties of moth-eye ARs gives rise to new directions to increase the PCE of PSCs. 300 nm and 1000 nm inverted moth-eye structured polydimethylsiloxane (PDMS) films were fabricated using soft lithography and reported in \cite{kim2019moth}. The former one abated the optical loss at the air/glass interface and enhance the solar cell efficiency by about 21\% from 19.66\% and $J_{sc}$ from 23.83 mA/cm$^2$ to 25.11 mA/cm$^2$. The 1000 nm moth-eye structured PDMS films exhibit elegant coloration due to the interference originating from the diffraction grating effect. Ormostamp-based moth-eye AR has been fabricated using the nano-imprinting method and included on the glass side of the all-inorganic carbon-based CsPbIBr$_2$ PSCs in \cite{lan2021performance}. This resulted in an increase of $J_{sc}$ from 10.89 mA/cm$^2$ to 11.91 mA/cm$^2$ and PCE increased from 9.17 \% to 10.08 \%. The solar cell also showed excellent adaptability to higher temperatures up to 200$^\circ$. \\

A superhydrophobic surface is another application of moth-eye structures apart from the anti-reflection coatings. Moth-eye textured surfaces, due to their roughness, possess enhanced water-repellent properties, resulting in a superhydrophobic surface. Such structures have been utilized to fabricate foldable displays for various electronic devices. The transmission property of the structures remains almost invariant under different bending, thermal, and chemical conditions. Hence, the resulting displays show very good anti-reflective function, excellent mechanical resilience, and foldability, as well as superhydrophobicity with good thermal and chemical resistance\cite{yun2020superhydrophobic}.

\subsection{Anti-reflective Transparent Nanostructures: }
\textit{Greta Oto}, commonly known as glasswing butterfly, and the cicada Cretensis species possesses unique transparent wings with broadband and omnidirectional anti-reflection property. This property originates from the presence of nanopillars having periodicity in the range of 150-250 nm on the wing's surface resulting in a gradient of refractive index between air and wing's membrane\cite{binetti2009natural, siddique2015role, morikawa2016nanostructured}. This property inspired the fabrication of omnidirectional anti-reflective glass which can show reflectivity smaller than 1\% irrespective of incident angles in the visible spectrum for S-P linearly polarized configurations\cite{papadopoulos2019biomimetic}. Such biomimetic anti-reflective transparent materials have a pervasive application including display, solar windows, optical components, and devices\cite{choi2016novel, gombert2000antireflective, yu2009efficiency, jannasch2012nanonewton, gao2017optical, xi2007optical, diao2016nanostructured, pfeiffer2017antireflection}. Single-step laser texturing approach, nanosecond, and femtosecond laser processing systems are used to fabricate such anti-reflective transparent structures\cite{papadopoulos2019biomimetic, lou2019design, livakas2020omnidirectional}. 
\section{Biomimetic Nanotechnology in Textile Engineering}
Textile has been an integral part of human civilization for a long time. Humans started using textiles to protect themselves from the cold and other adverse conditions. With time, the trend and manufacturing process in textile has amended tremendously. Natural elements showing superhydrophobicity, self-cleaning property, structural color, and fine quality of fabric have inspired textile engineers and researchers to develop more sophisticated textiles with enhanced performances and functionalities. In this section, we'll review some of the research work related to nano-biomimicry in the field of textile engineering. 
\begin{figure*}[h]
 \centering
 \includegraphics[width = \textwidth]{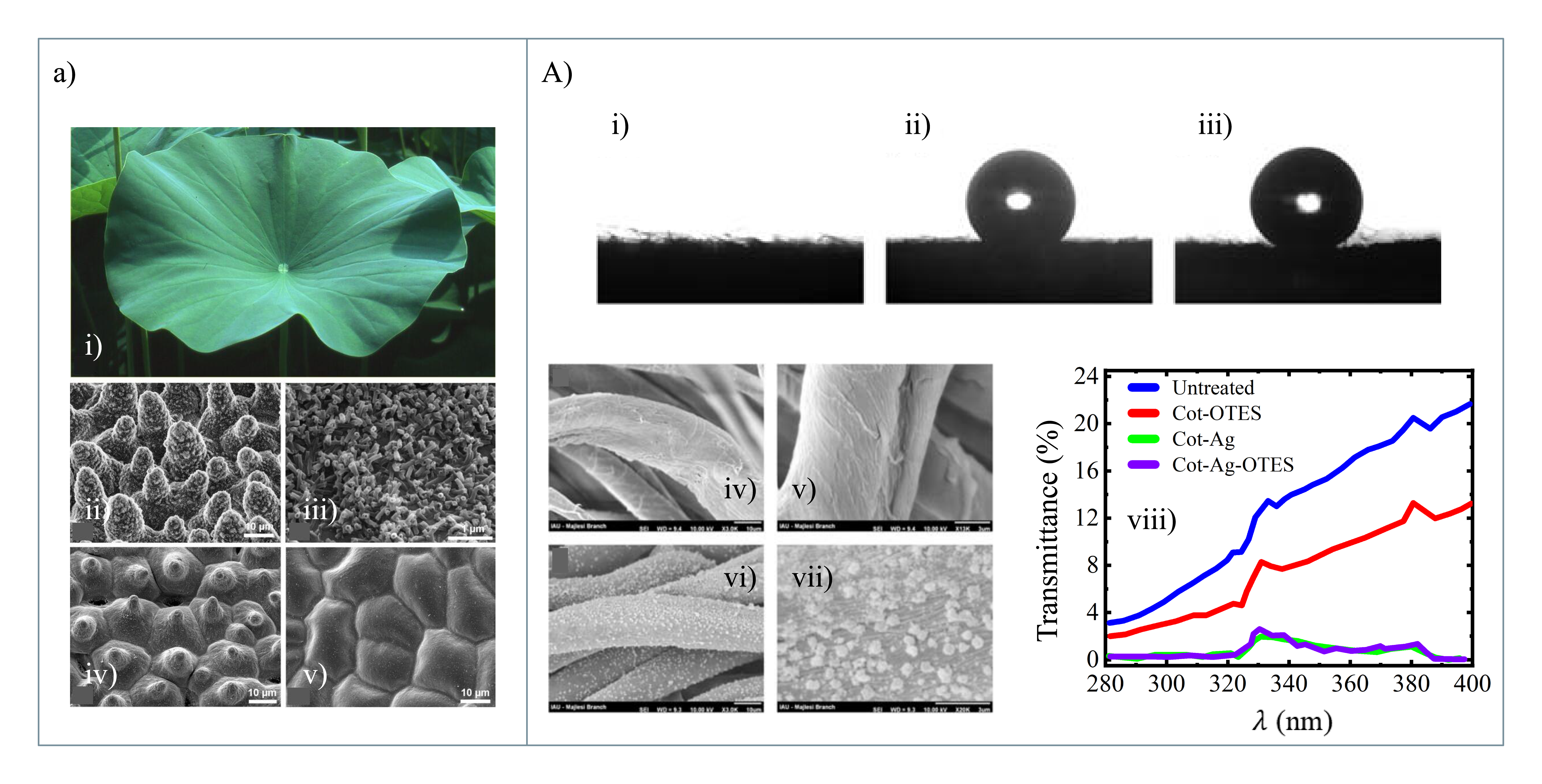}
 \caption{a) i) A natural super-hydrophobic surface Lotus Leaf ii) SEM image of the upper leaf side prepared by ‘glycerol substitution’ with the hierarchical surface structure consisting of papillae, wax clusters, and wax tubules. iii) Wax tubules on the upper leaf side. (iv) Upper leaf
side after critical-point (CP) drying and dissolution of wax tubules, which make the stomata visible. v) CP dried underside leaf shows convex cells without stomata.\cite{ensikat2011superhydrophobicity} A) A 5$\mu$L water droplet on the surface of i) untreated ii) the Cot-OTES iii) Cot-Ag-OTES. It is observed from these figures that with Ag coating the contact angle is greater than the other two and shows superhydrophobicity. This is inspired by the superhydrophobicity of the lotus leaf. SEM figure of iv)-v) untreated and vi)-vii) Cot-Ag samples. The water droplets are seen in the cases of vi) and vii). viii) UV-transmission spectra for different types of cotton samples. The UV blocking performance is better in the case of Ag-coated samples.\cite{shateri2013fabrication}(Images in this figure were reproduced from the cited original work, with permission from their corresponding publishers.)}
 \label{textile1}
\end{figure*}
\subsection{Water Repellent Textile}
Superhydrophobicity and self-cleaning properties are prevalent in nature in various plants. The most common and notable example is the lotus. Due to the higher contact angle ($\geq150^\circ$), virtually no wetting of the surface takes place on a superhydrophobic surface, leading to the property of self-cleaning. Whenever rain falls on a lotus leaf, the water droplets roll off as a result of a higher contact angle, and on its way, it collects the dirt or other material, cleaning the surface and keeping the leaves dry and free of pathogens\cite{marmur2004lotus, balani2009hydrophobicity, zhang2006superhydrophobic, das2017potential}. \\
\noindent Nanotex was founded in 1998. It was based on the idea of superhydrophobicity i.e., repelling water from a surface. They implemented this natural process using nanotechnology, where nano molecules bonded to the textiles provided an efficient stain-repellent mechanism. This eventually led to the removal of stains and dirt. This technology is now used in over 100 brands around the world. In another research work published in 2013, the superhydrophobicity of lotus leaves inspired the development of a simple coating method using silver nanoparticles (AgNPs). This could facilitate the bionic creation of superhydrophobic surfaces on cotton textiles with the functionality of water repellency, antibacterial, and UV blocking\cite{shateri2013fabrication}.\\

In this work, NaOH-soaked cotton fiber was immersed into the aqueous solution which resulted in the formation of Silver cellulosate which finally formed AgNPs through reduction of Ag cellulosate by $C_6H_6O_8$ solution. This Ag-coated cotton fabric was then immersed into the prehydrolyzed OTES (octyltriethoxysilane) and the final fabric was called "Cot-Ag-OTES". This fabric was then subjected to ultraviolet-visible reflectance spectrophotometry, scanning electron microscopy, energy dispersive X-ray spectroscopy, and X-ray diffractometry for characterization. The superhydrophobicity was determined by measuring a 5$\mu$L water droplet contact angle (CA) and shedding angle (SHA). Cot-OTES (The sample without AgNPs) had a CA of $135^\circ \pm 5.2^\circ$ and water SHA of $31^\circ$. But with the introduction of AgNps coating, the CA was found to be $156^\circ \pm 3.8^\circ$ and SHA of $8^\circ$. This proves the performance enhancement of the water repellence property of the Cot-Ag-OTES sample. 

Figure \ref{textile1}(A)(i-iii) show the optical micrographs of 5$\mu$L water droplets on the surfaces of the untreated (i) Cot-OTES (ii) Cot-Ag-OTES (iii) fabric samples. With the introduction of AgNps, the increase of contact angle and hence the hydrophobic property is evident in this figure. SEM micrographs of untreated and Agnps treated cotton samples are given in figure \ref{textile1}(A)(iv-v) and (A)(vi-vii) respectively. In \ref{textile1}(A)(vii), the water droplets are visible on the sample, ensuring the hydrophobic property. Gram-negative E.coli and gram-positive S. aureus bacteria were used to evaluate the anti-bacterial property of the samples. It turned out that the inhibition zone against E.coli and S.aureus was about 2 and 3nm respectively which indicates high antibacterial activity.\\

\noindent Transmittance spectra (shown in figure \ref{textile1}(A)(viii) of the samples in the UV range (280-400nm) were also observed to assess the UV-blocking property. From this figure, it is pretty clear that the transmittance is very low for AgNPs-coated fabrics with and without the OTES layer, which manifests itself as excellent at blocking UV. The sample Cot-Ag-OTES happened to have a UPF(Ultraviolet Protection Factor) value of 266.01 where UPF$>$40 is considered to be excellent against UV radiation. So the fabric produced in this work with Silver Nanoparticles coating on cotton mimicking the lotus leaf showed excellent performance as superhydrophobic, anti-bacterial, and UV-blocking fabric.\\

Bashari et al. proposed a natural element-based water repellent nano-coating instead of using C8 fluorochemicals compounds onto different fabrics using a layer by layer self assembly\cite{bashari2020bioinspired}. To mimic the superhydrophobicity observed in nature, various water-repellent finishing agents are used. Among them, fluorocarbons (FCs) possess a unique property of repelling not only water but also, oil providing a surface tension of $10-20 mNm^{-1}$ while coated on a fabric. However, it has been reported to exert a detrimental effect on the environment and humans leading to the need for alternative green solutions. Carnauba wax obtained from a Brazilian tree carnauba palm, formally known as \textit{Copernicia Prunifera}, is a natural water repellent agent which has been used in this proposed method to develop hydrophobic textiles using layer-by-layer self-assembly. This natural water-repellent nano-coating was deposited on three types of samples, cotton, nylon, and cotton-nylon fabrics. \\

Characterization of the solid carnauba nanoparticles was done through various measurements and experiments. Table \ref{table2_textile} shows the contact angles and the antibacterial effects of the untreated and treated samples. It is seen that all of the treated samples have significant hydrophobicity for 30 sec and even after washing. The coated fabrics show enhanced antibacterial effects against \textit{Staphylococcus aureus} (\textit{S.aureus}) and \textit{Escherichia coli} (\textit{E.coli}). The coated fabrics also show good air permeability and thus have paved the way for green ingredient-based environmentally friendly water repellent textiles.
\begin{table*}[hbt!]
\centering
\caption{Contact angle and Antibacterial Effect of different samples with water as the wetting medium.}
\label{table2_textile}
{\begin{tabular}{lllllll}
\multirow{2}{*}{Sample}  & \multirow{2}{*}{Treatment} & \multicolumn{3}{l}{\begin{tabular}[c]{@{}l@{}}Averaged Contact \\angle (degree)\end{tabular}} & \multicolumn{2}{l}{Antibacterial Effect (\%)}  \\ 
\cline{3-7}
                         &                            & 0s    &  & 30s                                                                                & S. aureus & E.coli                             \\ 
\hline
\multirow{3}{*}{Cotton}  & Untreated                  & 45.36 &  & 0                                                                                  & -         & -                                  \\
                         & Treated                    & 127.6 &  & 130.9                                                                              & 95.75     & 95.1                               \\
                         & Treated~ After Washing     & 109.9 &  & 101.7                                                                              &     -      &    -                                \\ 
\hline
\multirow{3}{*}{Nyco}    & Untreated                  & 141.6 &  & 48.7                                                                               &    -       &     -                               \\
                         & Treated                    & 132.1 &  & 135                                                                                & 95.75     & 94.6                               \\
                         & Treated After Washing      & 122.8 &  & 124.9                                                                              &    -       &            -                        \\ 
\hline
\multirow{3}{*}{Nylon 6} & Untreated                  & 128.6 &  & 77.3                                                                               &    -       &    -                                \\
                         & Treated                    & 127   &  & 131.4                                                                              & 97.78     & 94.1                               \\
                         & Treated After Washing      & 105.2 &  & 102.4                                                                              &   -        &          -                          \\
\end{tabular}
}
\end{table*}

\subsection{Textile dyeing}
 Natural instances show a great variety of structural colors that inspired engineers and designers to design various dyeing techniques and processes. Biomass pigments, with their sustainability, naturality, and biocompatibility, have become an interesting research object in cleaner production-focused textile dyeing. Microbial pigments have the advantage of not being constrained by season, climate, or geography\cite{panesar2015production,kumar2015microbial,tuli2015microbial,nigam2016food, dufosse2014filamentous, venil2014current, venil2013bacterial}. Prodigiosin, an intracellular metabolite with a pyrrole structure, has bright colors and antibacterial function, making it attractive to the researchers\cite{falklof2016steric},\cite{liu2013mutant}.In the past pyrrole structure red pigment was insoluble in water and had to use organic solvents to extract from the interior of thalli which increased the cost\cite{alihosseini2008antibacterial},\cite{kim2015dyeing}. Now, if the pyrrole structure pigment is produced by cell metabolization, it will result in cost reduction and extraction of microbial pigments can be avoided\cite{gong2019bio}. To design this trans-membrane culture, it is required to mimic the mechanism and substance transport through the cell membrane. Through the simulation of the artificial membrane, the chemistry of the preparation and kinetic study of the actual cell membrane can be conducted. Phospholipid bilayers are dynamic and closely similar to cell membranes, and this can be used to research trans-membrane behavior. These phospholipid membrane microbial pigments can cross the membrane easily for smaller molecular structure media\cite{gong2019bio}. It was found in the work presented in \cite{gong2019bio} for positively charged substances that the pigments had a better mass transfer rate than negatively charged segments. The application of electric fields also boosted the extracellular substance concentration. Using a permeability agent also improved the permeability of pigments. 
\section{Conclusions}
In this review paper, we have summarized and highlighted several state-of-the art innovations resulted from the application of biomimicry in nanotechnology. We briefly illustrated how biomimicry can amend various nanotechnology based field such as nano-sensors, nano-robots, nano-energy harvest and so on. Bio-compatiable nano robots and sensors built taking inspirations from nature will play a crucial role in medical science and also in making human life better and more comfortable. Photonic structures mimicked from natural instances have paved the way for better performing photonic devices. Artificial nano leaves, kelp-inspired TENG and using biomimetic nanoparticles in solar cell can facilitate our energy harness capability. Using nanoparticles and natural water repellent on fabrics might replace the existing repellent agent and  will be less toxic and more eco-friendly. In a nutshell, application of biomimicry in nanotechnology can provide unique, eco-friendly and efficient solution to various problems faced by humans. We deem that the biologists will discover more astounding information in future and the nanotech researchers will exploit them to build a better future for human civilization.

\section*{Conflicts of interest}
The authors declare no conflicts of interest.

\section*{Acknowledgements}
The authors acknowledge the support of both the Department of Electrical and Electronic Engineering, Bangladesh University of Engineering and Technology, and the Department of Computer Science and Engineering, Brac Univerisity.


\balance


\bibliography{rsc} 
\bibliographystyle{rsc} 

\end{document}